\documentclass[12pt]{article}

\usepackage[dvips]{epsfig}
\usepackage[]{amsbsy}

\begin{document}

\title{Metallic surface of a Mott insulator - \\ 
       Mott insulating surface of a metal}

\author{M. Potthoff and W. Nolting}

\maketitle

\begin{center}
Humboldt-Universit\"at zu Berlin, Institut f\"ur Physik,
D-10115 Berlin, Germany
\end{center}

\vspace{0.5cm}

\begin{center}
\parbox{140mm}{ \small

{\bf Abstract} 

The dynamical mean-field theory (DMFT) is employed to study the
correlation-driven metal-insulator transition in the semi-infinite
Hubbard model at half-filling and zero temperature. We consider
the low-index surfaces of the three-dimensional simple-cubic
lattice and systematically vary the model parameters at the very
surface, the intra- and inter-layer surface hopping and the surface
Coulomb interaction. Within the DMFT the self-energy functional is
assumed to be local. Therewith, the problem is self-consistently 
mapped onto a set of coupled effective impurity models corresponding
to the inequivalent layers parallel to the surface. Assuming that
the influence of the high-energy Hubbard bands on the low-energy
quasi-particle resonance can be neglected at the critical point, 
a simplified ``linearized DMFT'' becomes possible. The linearized 
theory, however, is formally equivalent to the Weiss molecular-field 
theory for the semi-infinite Ising model. This implies that 
qualitatively the rich phenomenology of the Landau description 
of second-order phase transitions at surfaces has a direct analogue 
for the surface Mott transition. Motivated by this formal analogy, 
we work out the predictions of the linearized DMFT in detail. 
It is found that under certain circumstances the surface of a 
Mott insulator can be metallic while a Mott-insulating surface 
of a normal metal is not possible. We derive the corresponding
phase diagrams, the (mean-field) critical exponents and the critical
profiles of the quasi-particle weight. The results are confirmed by 
a fully numerical evaluation of the DMFT equations using the
exact-diagonalization (ED) method. By means of the ED approach we
especially investigate the non-critical parts of the phase diagrams
and discuss the $U$ and layer dependence of the quasi-particle
weight. For strong modifications of the surface model parameters, the
surface low-energy electronic structure dynamically decouples from 
the bulk.
}
\end{center}
\vspace{8mm}



\newpage

\section{Introduction} 
\label{sec:s1}

The correlation-driven transition from a paramagnetic metal to a
paramagnetic Mott-Hubbard insulator \cite{Mot49,Mot90} 
constitutes one of the fundamental problems in solid-state theory.
The Mott transition is interesting since strong electron correlations
lead to low-energy electronic properties that are qualitatively 
different from those predicted by band theory.

Since it has been recognized that the limit of infinite spatial 
dimensions ($D=\infty$) is a well-defined and meaningful limit also 
for itinerant-electron models \cite{MV89} and since the invention
of dynamical mean-field theory (DMFT) \cite{Vol93,GKKR96}, there is a 
renewed interest in the Mott transition \cite{Geb97}. The DMFT 
provides an in principle exact description of the transition 
in infinite dimensions. While this is a somewhat artificial limit, 
the DMFT, as a mean-field concept, represents a powerful approach 
also for the study of finite-dimensional systems. Analogous to the 
Weiss molecular-field theory for localized spin models, the DMFT 
can be expected to give a valuable mean-field picture of the physics 
of three-dimensional itinerant-electron models \cite{Eng63}.

Presumably, the simplest model that includes the essentials of the 
Mott transition is the Hubbard model \cite{Gut63,Hub63,Kan63}. From 
the application of the iterative perturbation theory (IPT) \cite{GK92a}
to the $D=\infty$ Hubbard model at half-filling and zero temperature, 
the following scenario for the Mott transition has emerged 
\cite{GKKR96}: For small Coulomb interaction $U$, the system is a 
metallic Fermi liquid with a quasi-particle peak at the Fermi energy 
in the one-electron spectrum. As $U$ approaches a critical 
value $U_{c2}$ from below, the quasi-particle weight vanishes 
continuously, similar as in the Brinkman-Rice approach \cite{BR70}.
For strong $U$ the system is a Mott-Hubbard insulator with an 
insulating gap in the one-electron spectrum, similar to the 
Hubbard-III approach \cite{Hub64b}. The insulating solution ceases 
to exist when $U$ approaches another critical value $U_{c1}$ from 
above. In the entire coexistence region $U_{c1} < U < U_{c2}$ the
metallic solution is stable, and thus the transition is of second
order. The pre-formed gap opens discontinuously at $U_c=U_{c2}$.

It has been questioned \cite{LN98,Noz98,Keh98,NG99} whether
the picture given by the IPT is correct. Recent numerical 
renormalization-group calculations (NRG) \cite{BHP98,Bul99}, however, 
corroborate the IPT results qualitatively, although $U_c$ is found 
to be significantly smaller than in the perturbational approach 
\cite{finitet}. On the other hand, there is a remarkable agreement 
of the NRG with the result for $U_c$ in the projective self-consistent 
method (PSCM) \cite{MSK+95}.

The NRG calculations show that for $U\mapsto U_c$ the quasi-particle
resonance becomes more or less isolated from the high-energy Hubbard 
bands \cite{Bul99}. The resonance basically reproduces itself in the 
self-consistent evaluation of the mean-field equations. This fact can
be used for a simplified treatment of the DMFT where
the influence of the Hubbard bands on the low-energy peak is neglected
\cite{BP99}. This ``linearized DMFT'' yields a simple algebraic 
equation for the quasi-particle weight at the critical interaction
and thereby allows for an analytical estimate of $U_c$. The results
are in good agreement with the numerical values for $U_c$ obtained
from NRG and PSCM on different lattices \cite{BP99}. For inhomogeneous
systems, the linearized DMFT also determines the critical profile of 
the quasi-particle weight and the dependence of $U_c$ on the system 
geometry. Comparing with the numerical results obtained from the exact 
diagonalization method (ED), a convincing qualitative agreement with 
respect to the thickness and geometry dependence of $U_c$ has been 
found for thin Hubbard films \cite{PN99a}.

It has been noticed \cite{PN99a} that the equation of the linearized
DMFT is of the same form as the linearized mean-field equation of the
Weiss molecular-field theory for the Ising model (at the critical
temperature). There is a one-to-one correspondence if one identifies 
the quasi-particle weight $z$ with the magnetization $m$, the squared 
interaction $U^2$ with the temperature $T$, and the squared hopping 
integral $t^2$ with the exchange coupling $J$:
\begin{equation}
z \Leftrightarrow m
\: , \;\;\;
U^2 \Leftrightarrow k_B T 
\: , \;\;\;
36t^2 \Leftrightarrow J/2
\: .
\label{eq:ident}
\end{equation}
The Weiss theory, 
on the other hand, can be considered as being a coarse-grained 
realization of the classical Landau theory of second-order phase 
transitions \cite{LL59}. Consequently, the results of Landau theory 
(for $T=T_c$) can be translated back into predictions concerning 
the Mott transition in the Hubbard model (for $U=U_c$). 

While the 
Landau theory of phase transitions is rather simple for homogeneous 
systems, the mean-field theory of critical behavior at {\em surfaces} 
is much more involved and numerous non-trivial results can be derived 
\cite{Bin83}. The idea of the present paper is thus to take the 
Landau theory as starting point and motivation to work out the 
predictions of the linearized DMFT for Hubbard surfaces and finally 
to test the predictions, as far as possible, by comparing with a 
fully numerical solution of the DMFT equations.

Within the classical Landau theory, the free energy is expanded in 
terms of the local order parameter $m({\bf r})$. For a semi-infinite 
system (surface geometry) one additionally considers a surface 
contribution to the free energy \cite{Bin83}. Laterally, the order 
parameter is assumed to be homogeneous. 
We take the $x$ axis be parallel 
to the surface normal and pointing into the volume ($x>0$), then 
$m=m(x)$, and $m(x=0)$ is the surface value of the order parameter.
Let us list those mean-field predictions derived from the 
Ginzburg-Landau free energy \cite{Bin83} which -- by means of the
above-mentioned formal analogy -- have a direct counterpart
for the Mott transition:

\begin{enumerate}
\item 
The transition in the bulk of the semi-infinite system occurs
exactly at the same critical temperature $T_{c,{\rm bulk}}$ as 
for the infinitely extended system: $T_{c,{\rm bulk}}=T_c$.
\item
Near the surface the order-parameter profile $m(x)$ vanishes at a 
distance $\Lambda$ beyond the surface if linearly extrapolated from 
the boundary. The so-called ``extrapolation length'' $\Lambda$ as 
well as the (bulk) correlation length $\xi$ are the two length scales 
that characterize the order-parameter profile in the continuum model.
Microscopically, the extrapolation length is related to the model 
parameters at the surface. In the molecular-field approximation of 
the Ising model we have $\Lambda^{-1} \propto (\Delta_c - \Delta)$ 
where $\Delta$ is the modification of the exchange coupling within 
the surface layer, $J_{11} = J (1 + \Delta)$, and $\Delta = \Delta_c$ 
corresponds to $\Lambda = \infty$.
\item
For uniform parameters ($\Delta=0$) the mean field is smaller at 
the surface due to missing neighbors. This implies a weaker tendency 
to order. $m(x=0)$ is smaller than $m(x\mapsto \infty) = m_{\rm bulk}$, 
and $m(x)$ monotonously increases with increasing $x$ (this implies 
$\Lambda > 0$). There is a finite order parameter $m(x=0)>0$ at the 
surface only for $T<T_c$, i.~e.\ only when there is spontaneous 
order also in the bulk. The transition at $T_c$ 
is termed the ``ordinary transition''.
\item
For $\Lambda < 0$ ($\Delta>\Delta_c$) the surface layer orders at 
a temperature $T_{c,{\rm surf}} > T_{c,{\rm bulk}}$ (``surface 
transition''). In the regime $T_{c,{\rm bulk}} < T < T_{c,{\rm surf}}$, 
the bulk correlation length $\xi$ is finite, and the order parameter 
decays exponentially fast from its maximum value $m(x=0)$ at the 
surface towards zero in the bulk. At $T=T_{c,{\rm bulk}}$ 
(``extraordinary transition''), the divergence of $\xi$ and the 
onset of order in the bulk induce singularities in the behavior of 
surface response functions. For the order 
parameter at the surface $m(x=0)$, there is a discontinuity of its
second derivative only. At $=T_{c,{\rm bulk}}$ the order-parameter
profile decays algebraically, $m(x) \propto 1/x$.
\item
It holds: $(T_{c,{\rm surf}} - T_{c,{\rm bulk}}) / 
T_{c,{\rm bulk}} \propto \Lambda^{-2}$. The transition at 
$T = T_{c,{\rm surf}} = T_{c,{\rm bulk}}$ in the case 
$\Lambda = \infty$ is called the ``special transition''.
For $\Lambda = \infty$ the order-parameter profile 
is flat in the ordered phase; the trivial solution
$m(x) = m_{\rm bulk} = {\rm const.}$ minimizes the 
Ginzburg-Landau free energy. In this situation the effect
of missing neighbors at the surface is compensated exactly.
The topology of the phase diagram
(ordinary, extraordinary, surface and special transition)
should be correct {\em whenever}
the surface can support independent order \cite{MW66}.
For example, there is no surface transition in the semi-infinite 
two-dimensional Ising model since the ``surface'' is 
one-dimensional \cite{AY73}.
\item
There are two critical exponents that are merely related to 
the critical temperatures (instead of describing the critical 
behavior of order parameter and response functions), the 
``shift exponent'' $\lambda_s$ and the ``crossover exponent''
$\phi$. They are defined as 
$(T_c(d)-T_c(d=\infty))/T_c(d=\infty) \propto d^{-\lambda_s}$ for
$d\mapsto \infty$ where $T_c(d)$ is the critical temperature of 
a film of finite thickness $d$, and
$(T_{c, {\rm surf}}(\Delta)-T_{c,{\rm bulk}}) / T_{c,{\rm bulk}} 
\propto (\Delta/\Delta_c-1)^{1/\phi}$ for $\Delta \mapsto \Delta_c$
(special transition).
Within Landau mean-field theory one has $\lambda_s=2$ and
$\phi=1/2$.
\item
Spontaneous order in the bulk always induces a finite order 
parameter at the surface, $m(x=0)>0$.
\end{enumerate}

The Landau theory also makes additional statements concerning e.~g.\
the bulk and surface critical exponents of the order parameter as well 
as the exponents of response functions with respect to an external 
applied field. We do not mention such results in the present context,
since either they have no obvious analogue for the Mott transition
(applied field) or they refer to temperatures $T \mapsto T_c$ but 
$T \ne T_c$ where the mean-field equation cannot be linearized and
where the formal analogy (\ref{eq:ident}) breaks down. We will, 
however, discuss a simple extension of the linearized DMFT for 
$U \mapsto U_c$ but $U \ne U_c$ which recovers the result 
$z \propto (U_c - U)$ of the PSCM \cite{MSK+95}.

To a certain extent, the phase diagram predicted by the Landau 
theory or, respectively, by the linearized DMFT can be tested by
comparing with a fully numerical evaluation of the DMFT equations. 
We need an approach that is sufficiently simple for a systematic study
of a large number of geometries and model parameters. For this purpose
the exact-diagonalization method of Caffarel and Krauth \cite{CK94}
is well suited. We mainly focus on the non-critical parts in the 
phase diagram where the ED is able to give reliable results. 
Critical exponents, for example, cannot be calculated reliably.
The ED has successfully been employed beforehand for the discussion 
of the Mott transition in thin Hubbard films \cite{PN99a} and 
at Hubbard surfaces \cite{PN99c} where the film and surface 
electronic structure has been discussed in detail. Contrary, the 
present paper focuses on the surface modification of the model 
parameters and on surface phases and thereby substantially extends 
the previous studies.

The Mott transition at a surface of the semi-infinite Hubbard model 
has recently been investigated in a paper of Hasegawa \cite{Has92}
on the basis of the Kotliar-Ruckenstein slave-boson theory \cite{KR86}.
With the present study we methodically improve upon Hasegawa's work.
We will also show that for $U \mapsto U_c$ the perturbation of the 
system that is introduced by the presence of the surface deeply 
extends into the bulk. It is thus insufficient to assume (local)
physical quantities to be different from their value in the bulk
only in the first few surface layers. Such a restriction gives rise 
to unphysical singularities, e.~g.\ in the $U$ dependence of the 
quasi-particle weight as they are seen in Ref.\ \cite{Has92}. Within 
the slave-boson theory it is found that under certain circumstances 
a metallic surface can coexist with an insulating bulk \cite{Has92}. 
Crucial for the existence of this surface phase is a considerable 
decrease of $U$ at the surface. This is an interesting and also 
plausible result although the required strong decrease of $U$ at 
the surface appears to be quite unrealistic for real systems.

A physically more relevant modification of the model parameters is, 
in first place, the enhancement or decrease of the hopping integrals 
at the surface. This may be caused by a relaxation of the interlayer 
distance, for example. According to the scaling law $t \sim r^{-5}$ 
for $d$ electrons (cf.\ e.~g.\ Ref.\ \cite{Pap86}), a top-layer 
relaxation $\Delta r/r$ of a few per cent results in a strong change 
of the hopping integral. 
A surface modification of $t$ up to about 10\%-20\% 
appears to be realistic. Besides the hopping we will also discuss a 
modification of $U$ at the surface. In 3d transition metals, however, 
this effect seems to be less important \cite{DDP95,PPP96}. In any case,
$U$ is expected to be enhanced at the surface \cite{PPP96}. 
On the contrary, it will be shown that the interesting surface phase 
occurs for {\em lowered} surface $U$. Another important aspect 
is the surface geometry which is expected to affect 
the surface phase diagram considerably. Open surfaces with a strong 
reduction of the surface coordination number will show the most 
pronounced surface effects in the electronic structure. We thus
consider different low-index surfaces of a $D=3$ simple-cubic 
lattice.

The basic assumption of DMFT is the strict locality of the 
self-energy functional. For $D=3$ dimensions this represents 
a strong simplification of the problem. The local approximation 
is well justified for the weak-coupling regime, also for the 
case of surface geometries (see the discussion in Refs.\ 
\cite{PN99c,SC91,PN97c}). For the intermediate- to strong-coupling 
regime, however, the assumption may be questioned. One could
alternatively investigate a surface of a $D=\infty$ lattice
where the DMFT becomes exact. While this will be discussed briefly, 
we otherwise consider surfaces in $D=3$ dimensions. As in Refs.\ 
\cite{PN99a,PN99c} we expect the mean-field concept to be a good 
starting point for $D=3$.

The plan of the paper is the following:
The next section introduces into the model. The application of
DMFT for surface geometries is briefly discussed in Sec.\ 
\ref{sec:s3}.
We use two different methods to solve the DMFT equations: The
first one is the approximative linearization of the equations for
$U=U_c$ \cite{BP99}. This is presented in Sec.\ \ref{sec:s4}. 
Sec.\ \ref{sec:s5} then 
gives a discussion of the analytical results. For the full 
solution of the DMFT equations, we employ the exact-diagonalization 
method \cite{CK94} which is introduced in Sec.\ \ref{sec:s6}. 
The corresponding
results are discussed in Sec.\ \ref{sec:s7}. 
Finally, Sec.\ \ref{sec:s8} concludes the 
paper.

\section{Semi-infinite Hubbard model}
\label{sec:s2}

We investigate the Hubbard model on a three-dimensional, simple-cubic
and semi-infinite lattice. The system is considered to be built up
by two-dimensional layers parallel to the surface. Accordingly, the
position vector to a particular site in the semi-infinite lattice
is written as ${\bf R}_{\rm site}={\bf r}_i + {\bf R}_\alpha$. Here
${\bf R}_\alpha$ stands for the coordinate origin in the layer 
$\alpha$, and the layer index runs from $\alpha=1$ (topmost surface 
layer) to infinity (bulk). ${\bf r}_i$ is the position vector 
with respect to a layer-dependent origin and runs over the sites 
within the layer. In this notation the Hamiltonian reads:
\begin{equation}
  H = \sum_{ij \alpha \beta \sigma} t_{i\alpha,j\beta} \:
  c^\dagger_{i\alpha\sigma} c_{j\beta\sigma}
  + \sum_{i\alpha\sigma} \frac{U_\alpha}{2} \: 
  n_{i\alpha\sigma} n_{i\alpha -\sigma} \: .
\label{eq:hubbard}
\end{equation}
$\sigma=\uparrow, \downarrow$ is the spin index. $U_\alpha$ is 
the (layer-dependent) Hubbard interaction strength. The hopping 
integrals are restricted to be non-zero between nearest neighbors.
The energy zero is defined by setting $t_{i\alpha,i\alpha} 
\equiv t_0=0$ 
for sites in the bulk ($\alpha \mapsto \infty$). The energy scale 
is given by taking the (bulk) nearest-neighbor hopping to be
$t_{\langle i\alpha,j\beta \rangle}=-t$ with $t=1$.

The presence of the surface implies a breakdown of translational
symmetry with respect to the surface normal direction. Lateral
translational symmetry, however, may be exploited by performing
a two-dimensional Fourier transformation:
\begin{equation}
  \epsilon_{\alpha \beta} ({\bf k}) = \frac{1}{N_\|} 
  \sum_{ij} e^{-i {\bf k} ({\bf r}_i-{\bf r}_j)} \:
  t_{i\alpha,j\beta} \: .
\end{equation}
Here ${\bf k}$ is a two-dimensional wave vector of the first
surface Brillouin zone (SBZ), and $N_\|$ denotes the number 
of sites within each layer ($N_\| \mapsto \infty$). Let us 
briefly discuss the Fourier-transformed hopping matrix which 
reads:
\begin{equation}
  \boldsymbol{\epsilon}({\bf k}) = \left(
  \begin{array}{ccccc}
  t_{11} \, \epsilon_\|({\bf k}) + \Delta t_0   
                 & t_{12} \, \epsilon_\perp({\bf k}) &        
                 &        &
\\
  t_{21} \, \epsilon_\perp({\bf k})             
                 & t \, \epsilon_\|({\bf k})        
                 & t \, \epsilon_\perp({\bf k})      
                 &        &
\\
                 & t \, \epsilon_\perp({\bf k})      
                 & t \, \epsilon_\|({\bf k})    
                 & ..     &
\\
                 &   & .. & .. &\\
  \end{array} \right) \: .
\label{eq:hopmat}
\end{equation}
For $\alpha \ge 2$ the intra-layer (parallel) hopping and
the inter-layer (perpendicular) hopping are written as 
$\epsilon_{\alpha \alpha}({\bf k})=t\epsilon_\|({\bf k})$ and
$\epsilon_{\alpha \alpha+1}({\bf k})=t\epsilon_\perp({\bf k})$,
respectively. We consider three different low-index surfaces
of the sc lattice. The hopping matrix for the sc(100) surface 
is obtained from:
\begin{eqnarray}
  && \epsilon_\|({\bf k}) = - 2 (\cos(k_x) + \cos(k_y)) \: ,
  \nonumber \\ &&
  |\epsilon_\perp({\bf k})|^2 = 1 \: .
\label{eq:epsilon100}
\end{eqnarray}
The perpendicular hopping is ${\bf k}$-independent in this case.
For the (110) surface we have:
\begin{eqnarray}
  && \epsilon_\|({\bf k}) = - 2 \cos(k_x) \: ,
  \nonumber \\ &&
  |\epsilon_\perp({\bf k})|^2 = 2 + 2 \cos(\sqrt{2} k_y) \: ,
\label{eq:epsilon110}
\end{eqnarray}
and for the sc(111) surface:
\begin{eqnarray}
  && \epsilon_\|({\bf k}) = 0 \: ,
  \nonumber \\ &&
  |\epsilon_\perp({\bf k})|^2 = 3 + 
  2 \cos(\sqrt{2} k_y) 
  \nonumber \\ && \mbox{} \hspace{14mm}
  + 4 \cos(\sqrt{3/2} k_x) \cos(\sqrt{1/2} k_y) \: .
\label{eq:epsilon111}
\end{eqnarray}
Since two nearest neighbors are always located in two different 
(adjacent) layer, the intra-layer hopping vanishes in the last
case. Note, that only the absolute square of 
$\epsilon_\perp({\bf k})$
enters the physical quantities we are interested in.

At the very surface of the semi-infinite system, we consider 
three different possible modifications of the model parameters. 
Firstly, the intra-layer hopping within the topmost surface layer 
$t_{11}$ may differ from its bulk value (see Eq.\ (\ref{eq:hopmat})). 
Secondly, we allow for an altered hopping $t_{12} = t_{21} \ne t$ 
between the topmost and the sub-surface layer. Finally, the on-site 
Coulomb interaction strength is assumed to be layer-independent 
$U_\alpha=U=\mbox{const.}$ except for the topmost layer,
$U_{\alpha=1} \ne U$.

We restrict ourselves to the case of manifest particle-hole 
symmetry, namely a bipartite (sc) lattice, nearest-neighbor 
hopping and half-filling
($n = 2\langle n_{i\alpha\sigma} \rangle = 1$). In this case
the Fermi energy is given by $\mu = t_0 + U/2$. It is fixed by 
the bulk values for the on-site hopping and for the Hubbard 
interaction. Consider the atomic limit $t=0$ for a moment:
The positions of the two Hubbard ``bands'' in the bulk spectrum 
are given by: $E_{\rm low} = t_0-\mu$ and $E_{\rm high}=t_0-\mu +U$,
i.~e.\ they lie {\em symmetric} with respect to $\mu$. In thermal 
equilibrium $\mu$ is also the Fermi energy for the top layer. 
The Hubbard peaks in the surface density of states lie at $E_{\rm 
low}=t_0-\mu +\Delta t_0$ and $E_{\rm high}=t_0+\Delta t_0-\mu +U_1$,
where we have taken into account the top-layer modification of the 
interaction strength and where we have introduced an additional 
modification $\Delta t_0$ of the atomic level for top-layer sites
(see Eq.\ (\ref{eq:hopmat})). To maintain manifest particle-hole 
symmetry and to ensure $\langle n_{i\alpha\sigma} \rangle = 0.5$ 
for $\alpha=1$, the Hubbard peaks must again lie symmetric with 
respect to $\mu$. Thus, we need:
\begin{equation}
  \Delta t_0 = (U-U_1)/2 \: .
\label{eq:deltat0}
\end{equation}
With this choice for the top-layer on-site hopping, the local 
density of states (DOS) 
$\rho_\alpha(E)=(-1/\pi)\, \mbox{Im}\, \langle 
\langle c_{i \alpha \sigma};c^\dagger_{i \alpha \sigma} \rangle 
\rangle_E$ is a symmetric function of energy for each 
$\alpha$.

We finally introduce the intra- and inter-layer 
coordination numbers $q$ and $p$ which denote the number of 
nearest neighbors within the same layer and in one of the two
adjacent layers, respectively. We have:
\begin{eqnarray}
  && q=4 \: , \;\; p=1 \;\;\;\; \mbox{for sc(100)} \nonumber \\
  && q=2 \: , \;\; p=2 \;\;\;\; \mbox{for sc(110)} \nonumber \\
  && q=0 \: , \;\; p=3 \;\;\;\; \mbox{for sc(111)} \: .
\end{eqnarray}
The bulk coordination number is $Z=q+2p$.
The surface coordination number is $Z_{\rm S}=q+p$.

\section{Dynamical mean-field theory for surface geometries} 
\label{sec:s3}

The one-particle Green function $\langle \langle c_{i \alpha \sigma};
c^\dagger_{j \beta \sigma} \rangle \rangle$ contains any important 
information we are interested in. Its diagonal elements 
$G_\alpha(E) \equiv G_{i\alpha,i\alpha}(E) \equiv \langle 
\langle c_{i \alpha \sigma};c^\dagger_{i \alpha \sigma} \rangle 
\rangle_E$ can be written in terms of the hopping matrix 
(\ref{eq:hopmat}) and the self-energy:
\begin{equation}
  G_{\alpha}(E) =
  \frac{1}{N_\|} \sum_{\bf k}  \left(
  \frac{\bf 1}{(E+\mu) {\bf 1} - {\boldsymbol{\epsilon}({\bf k}) 
  - \boldsymbol{\Sigma}(E)}} 
  \right)_{\alpha \alpha} 
  \: .
\label{eq:onsitegf}
\end{equation}
The self-energy matrix is taken to be ${\bf k}$-independent and
diagonal, $\Sigma_{\alpha \beta}(E) = \delta_{\alpha \beta}
\Sigma_\alpha(E)$, with layer-de\-pen\-dent elements: We assume
that the self-energy is a strictly {\em local} quantity.

In the case of an infinitely extended lattice with full translational 
symmetry, this basic assumption leads to the well-known 
equations of dynamical mean-field theory \cite{Vol93,GKKR96}
which self-consistently map the bulk lattice problem onto a
single-impurity problem \cite{GK92a,Jar92}. The present case 
of reduced translational symmetry can be treated analogously:
A local self-energy implies that the Luttinger-Ward functional
\cite{LW60} depends on the local (but layer-dependent) propagators 
only: $\Phi = \Phi[\dots,G_\alpha(E),\dots]$. This in turn means 
that the self-energy of the $\alpha$-th layer is solely a functional 
of the local propagator: $\Sigma_\alpha(E) = 
\delta \Phi / \delta G_\alpha(E) = {\cal S}[G_{\alpha}(E)]$. The 
functional ${\cal S}$ is the same as in the case of an impurity 
problem, e.~g.\ the single-impurity Anderson model (SIAM), 
$\Sigma_{\rm imp}(E) = {\cal S}[G_{\rm imp}(E)]$, because the same 
type of skeleton diagrams occur in the expansion of the impurity
self-energy $\Sigma_{\rm imp}(E)$. With each layer $\alpha=1,2,...$ 
we therefore associate a SIAM,
\begin{eqnarray}
  H^{(\alpha)}_{\rm imp} 
  & \!\!\! = \!\!\! & \sum_\sigma 
  \epsilon_d^{(\alpha)} c^\dagger_\sigma c_\sigma
  + U_\alpha n_\uparrow n_{\downarrow}
  + \sum_{k\sigma} \epsilon_k^{(\alpha)} 
  a^\dagger_{k\sigma} a_{k\sigma}  
  \nonumber \\ 
  & \!\!\! + \!\!\! & \sum_{k \sigma} \left( V_k^{(\alpha)} 
  a^\dagger_{k\sigma} c_\sigma + \mbox{H.c.} \right) \: ,
\label{eq:siam}
\end{eqnarray}
with $\epsilon_d^{(\alpha)}=t_{i\alpha, i\alpha}$ and where the
conduction-band energies $\epsilon_k^{(\alpha)}$ and hybridization
strengths $V_k^{(\alpha)}$ chosen such that we have 
\begin{equation}
  \Delta^{(\alpha)}(E)=E+\mu-\epsilon_d^{(\alpha)}
  -\Sigma_{\rm imp}^{(\alpha)}(E) - G_\alpha(E)^{-1}
\label{eq:dmftsc}
\end{equation}
for the hybridization function $\Delta^{(\alpha)}(E-\mu)\equiv\sum_k 
(V_k^{(\alpha)})^2/(E-\epsilon_k^{(\alpha)})$. (Eq.\ (\ref{eq:dmftsc})
only provides an implicit definition of the hybridization function 
since $\Sigma_{\rm imp}^{(\alpha)}$ depends on $\Delta^{(\alpha)}$).
This implies at once the equality between the impurity Green
function of the $\alpha$-th SIAM, $G_{\rm imp}^{(\alpha)}(E)$, and the 
on-site lattice Green function in the $\alpha$-th layer $G_\alpha(E)$
and thus the equality between the respective self-energies,
$\Sigma_{\rm imp}^{(\alpha)}(E) = {\cal S}[G^{(\alpha)}_{\rm imp}(E)]$ 
and $\Sigma_\alpha(E) = {\cal S}[G_\alpha(E)]$.

The following iterative procedure then allows to solve the
semi-infinite Hubbard model within the dynamical
mean-field approximation: 
Starting from a guess for the layer-dependent self-energies 
$\Sigma_\alpha(E)$, we calculate the on-site Green function of
the $\alpha$-th layer using Eq.\ (\ref{eq:onsitegf}). Via Eq.\ 
(\ref{eq:dmftsc}), $G_\alpha(E)$ and 
$\Sigma_\alpha(E)=\Sigma_{\rm imp}^{(\alpha)}(E)$ determine
the hybridization function $\Delta^{(\alpha)}(E)$ of the $\alpha$-th
SIAM. The crucial step consists in solving the impurity models
for $\alpha=1,2,...$ to get the impurity self-energies
$\Sigma_{\rm imp}^{(\alpha)}(E)$ which are required for the next
cycle. The cycles have to be repeated until self-consistency is 
achieved.

Applying the DMFT to the semi-infinite Hubbard model means 
to map the original lattice problem onto an infinite set of 
impurity problems. The mapping is mediated by the self-consistency 
equation (\ref{eq:dmftsc}) for $\alpha=1,2,...,\infty$. For a 
given set of hybridization functions, each impurity model can 
be treated separately. There is, however, an indirect coupling 
via Eq.\ (\ref{eq:onsitegf}) in the self-consistency cycle 
since the on-site Green function of a given layer depends on 
all layer-dependent self-energies. The essential difference 
with respect to the case of an infinitely extended lattice with 
full translational symmetry where only one single-impurity model
and only one self-consistency condition is needed, consists
in the fact that for a semi-infinite system the sites within 
different layers have to be considered as non-equivalent.

\section{Linearized DMFT at the critical interaction} 
\label{sec:s4}

The zero-temperature Mott transition from a paramagnetic metal to a 
paramagnetic insulator is actually hidden due to antiferromagnetic 
order which is realized in the true ground state. To study the 
Mott transition, the solutions of the mean-field equations have 
to be enforced to be spin symmetric.
There have been numerous DMFT studies of the $T=0$ Mott transition 
in the recent past using different methods to solve the impurity 
problem: the iterative perturbation theory (IPT) 
\cite{GK92a,GK93,RKZ94}, the exact-diagonalization approach (ED) 
\cite{CK94,SRKR94,RMK94}, the projective self-consistent method 
\cite{MSK+95} as well as numerical renormalization-group calculations 
(NRG) \cite{BHP98,Bul99}. 

The IPT and, in first place, the NRG results show that for 
$U \mapsto U_c$ the quasi-particle resonance at $E=0$ is more 
or less isolated from the high-energy Hubbard peaks at 
$E \approx \pm U/2$. The resonance basically reproduces 
itself in the self-consistent procedure to solve the DMFT 
equations. A reasonable 
assumption is therefore that for $U = U_c$ the low-energy part 
of the SIAM hybridization function $\Delta^{(\alpha)}(E)$ consists 
of a single pole at $E=0$ only,
\begin{equation}
  \Delta^{(\alpha)}(E) \mapsto \frac{\Delta_N^{(\alpha)}}{E} \: ,
\label{eq:delta}
\end{equation}
and that the effect of the Hubbard bands can be disregarded 
completely. With this assumption a simplified, ``linearized'' DMFT
becomes possible \cite{BP99,PN99a}. There is an attractive 
feature of this method which outweights the necessity for a further
approximation: It allows for a fully analytic treatment of the 
mean-field equations, and an analytical expression for $U_c$ 
is obtained. Studying the dependencies of $U_c$ on the model
parameters can provide a valuable first insight into the problem.
The predictions of the linearized theory have been compared 
beforehand with fully numerical DMFT results for the Bethe and the
hypercubic lattice in $D=\infty$ \cite{BP99} and for the case of thin 
Hubbard films \cite{PN99a}. A satisfactory quantitative agreement 
has been 
noticed. This makes us confident that at least the correct
trends can also be predicted for the case of a semi-infinite 
lattice.

The details of the method can be found in Ref.\ \cite{BP99};
here we simply repeat the main idea and the final result: In 
the ansatz for the hybridization function (\ref{eq:delta}), 
$\Delta_N^{(\alpha)}$ denotes the layer-dependent coefficient 
in the $N$-th step of the self-consistency cycle. The aim is to
calculate $\Delta_{N+1}^{(\alpha)}$. The one-pole structure of 
the hybridization function corresponds to a well-defined SIAM 
with $n_s=2$ sites which can analytically be 
solved for each $\alpha$. In the 
one-particle excitation spectrum of the $\alpha$-th SIAM there 
are two $\delta$-peaks at $E \approx \pm U_\alpha/2$ as well as 
two $\delta$-peaks near $E=0$ corresponding to the (infinitely sharp) 
Kondo resonance for $U = U_c$ in the infinite ($n_s=\infty$) system. 
The layer-dependent weight 
of the resonance $z_\alpha$ can be read off from the solution. 
$z_\alpha$ determines the self-energy $\Sigma_\alpha(E)=U_\alpha/2 
+ (1-z_\alpha^{-1}) E + \cdots$ and via Eq.\ (\ref{eq:onsitegf}) 
the on-site Green function of the $\alpha$-th layer at low energies.
Using these results in the self-consistency equation 
(\ref{eq:dmftsc}) and insisting on the one-pole structure of 
the hybridization function, yields a new coefficient 
$\Delta_{N+1}^{(\alpha)}$. At this point the possible influence 
of the Hubbard bands is ignored. The final equation that 
relates $\Delta_{N+1}^{(\alpha)}$ to $\Delta_{N}^{(\alpha)}$
reads:
\begin{equation}
  \Delta^{(\alpha)}_{N+1} = \sum_\beta K_{\alpha \beta} \:
  \Delta^{(\beta)}_{N} \: ,
\label{eq:lindmft}
\end{equation}
where we have defined the following semi-infinite tridiagonal matrix:
\begin{equation}
  {\bf K} = 36 \left( \!\!\!
  \begin{array}{ccccc}
  q\: t_{11}^2 / U_1^2 & p\: t_{12}t / U_1U &               &  &\\
  p\: t_{12}t / U_1U   & q\: t^2 / U^2      & p\: t^2 / U^2 &  &\\
                       & p\: t^2 / U^2      & q\: t^2 / U^2 &..&\\
                       &                    & ..            &..&\\
  \end{array} \!\!\! \right) \: .
\label{eq:kmatrix}
\end{equation}

A self-consistent solution of the linearized mean-field equation
(\ref{eq:lindmft}) is given by a non-trivial fixed point of 
${\bf K}$. Let $\lambda_r$ denote the eigenvalues of ${\bf K}$. 
We can distinguish between two cases: If $|\lambda_r| < 1$ for 
all $r$, there is the trivial solution 
$\lim_{N\mapsto \infty} \Delta^{(\alpha)}_{N}
= 0$ only. This situation corresponds to the insulating solution
beyond the critical point. 
Contrary, if there is at least one $\lambda_r>1$,
$\Delta^{(\alpha)}_{N}$ diverges exponentially as $N\mapsto \infty$.
This indicates the breakdown of the one-pole model for the 
hybridization function in the metallic solution below the critical
point. The maximum eigenvalue thus determines via
\begin{equation}
  \lambda_{\rm max} = 
  \lambda_{\rm max}(q,p,U,t_{11},t_{12},U_1) = 1
\label{eq:cond}
\end{equation}
the critical model parameters.

At the critical point the mean-field equation (\ref{eq:lindmft}) 
can be written as $z_\alpha = \sum_\beta K_{\alpha\beta} z_\beta$
since the layer-dependent quasi-particle weight 
$z_\alpha \propto \Delta^{(\alpha)}$. Formally, this equation
can be compared with the Weiss mean-field equation for the layer
magnetizations $m_\alpha$ in the semi-infinite Ising model with 
coupling constant $J$. The linearized mean-field equation for 
$T=T_C$ reads: 
$m_\alpha=(J/2 k_B T)(q m_\alpha + p m_{\alpha+1} + p m_{\alpha-1})$ 
(we assume the model parameters at the surface to be unmodified for
the moment). The formal analogy with Eqs.\ (\ref{eq:lindmft}) and 
(\ref{eq:kmatrix}) is obvious and justifies the identification made 
in Eq.\ (\ref{eq:ident}) and the corresponding discussion in 
Sec.\ \ref{sec:s1}.

\section{Analytical results} 
\label{sec:s5}

From the basic equation (\ref{eq:cond}) we can calculate the
critical parameters for different cases.
First, we consider a system that is built up by a finite number of 
$d$ layers (film geometry). The model parameters are taken to be
{\em uniform}, i.~e.\ $t_{11} = t_{12} = t$ and $U_1 = U$ (at both 
surfaces). The eigenvalues of the $d$-dimensional 
matrix (\ref{eq:kmatrix}) can be 
calculated analytically for this case \cite{Ste90}:
\begin{equation}
  \lambda_r = \frac{36 t^2}{U^2} \left(
  q + 2 p \cos\left(\frac{r\pi}{d+1}\right)
  \right) \: ,
\label{eq:eigen}
\end{equation}
with $r=1,...,d$. Taking the maximum eigenvalue and solving 
for $U$ yields the thickness dependence of the critical interaction
\cite{PN99a}:
\begin{equation}
  U_{c}(d) 
  = 6 t \: \sqrt{q+2p \cos\left(\frac{\pi}{d+1}\right)} \: .
\label{eq:ucfilm}
\end{equation}
Expanding the result for $U_c$ in the limit of $d\mapsto \infty$
yields:
\begin{equation}
  \frac{U_c(d)-U_c(\infty)}{U_c(\infty)} 
  \propto d^{-\lambda_s}
\label{eq:ucfilminf}
\end{equation}
with a ``shift exponent'' $\lambda_s=2$. 

In the limit $d=\infty$ any of the two film surfaces represents a
semi-infinite system. From (\ref{eq:ucfilm}) we get for the critical 
interaction:
\begin{equation}
  U_{c,{\rm bulk}}
  = 6 t \: \sqrt{q+2p} = 6t \: \sqrt{Z} \: ,
\label{eq:ucbulk}
\end{equation}
which is the same result as is found when applying the method to 
the infinitely extended bulk system directly \cite{BP99}. We notice 
that for the case of uniform model parameters the linearized DMFT 
yields a unique critical interaction for the semi-infinite system 
which is the same as the bulk value. No surface phase is found. 
This observation is fully consistent with what has been obtained 
in previous numerical DMFT studies of the Mott transition at Hubbard 
surfaces \cite{PN99c} for {\em uniform} parameters. Despite the fact 
that at the surface the electronic structure has turned out to be 
modified considerably, a surface critical interaction different from 
the bulk value has not been found.

In the following we thus concentrate on a
semi-infinite system with {\em modified} parameters at the 
very surface. Also in this case the condition 
(\ref{eq:cond}) can be treated analytically:
To simplify the notation let us write
$K_{11}=a'$, $K_{12}=K_{21}=b'$ and $K_{\alpha\alpha}=a$,
$K_{\alpha\alpha\pm 1}=b$ for $\alpha \ge 2$ ($a,b,a',b'\ge 0$).
Let ${\bf K}(n)$ be the matrix that is obtained from the
semi-infinite matrix
${\bf K}={\bf K}(0)$ 
by deleting its first $n$ rows and columns. 
Furthermore, we define 
${\cal G}_n({\lambda}) \equiv \det(\lambda {\bf 1} - {\bf K}(n+1))
/ \det(\lambda {\bf 1} - {\bf K}(n))$. 
${\cal G}_n({\lambda})$ is the (1,1) or ``surface'' element of the 
Green matrix $(\lambda {\bf 1} - {\bf K}(n))^{-1}$.
Expanding the determinant $\det(\lambda {\bf 1} - {\bf K}(n+1))$ 
with respect to the upper left element, one easily 
verifies the recurrence relation 
${\cal G}_n({\lambda})^{-1} = \lambda - a - b^2
{\cal G}_{n+1}({\lambda})$ for $n\ge 1$.
However, all the ${\cal G}_n({\lambda})$
for $n\ge 1$ must be equal since the (off-)diagonal 
elements of ${\bf K}(n \ge 1)$ are constant. 
This results in a quadratic equation for ${\cal G}$, 
the solution of which is 
given by:
\begin{equation}
  {\cal G}(\lambda) = {\cal G}_{n\ge 1}(\lambda) =
  \frac{1}{2b^2} \left(\lambda - a \mp \sqrt{(\lambda -a)^2 - 4b^2}
  \right) 
\label{eq:gbulk}
\end{equation}
for $\pm (\lambda-a) > 0$. 
The eigenvalue spectrum of the semi-infinite matrix ${\bf K}$ 
consists of a continuous bulk part which can be read off from 
Eq.\ (\ref{eq:eigen}) for $d\mapsto \infty$ to be given by:
\begin{equation}
  |\lambda - a| \le 2b \: .
\label{eq:cont}
\end{equation}
This is just the region where $\mbox{Im}\, {\cal G}(\lambda)\ne 0$.
The largest eigenvalue in the bulk continuum is given by
$\lambda = a + 2b$ corresponding to the bulk critical interaction
given in (\ref{eq:ucbulk}). At $U=U_{c,{\rm bulk}}$ the bulk
undergoes the metal-insulator transition irrespective of the
state of the surface.

Under certain circumstances an additional discrete (``surface'') 
eigenvalue $\lambda_s$ may split off the bulk continuum. If a 
discrete eigenvalue exists, we must have 
${\cal G}_0({\lambda_s})^{-1}=0$. Using the result (\ref{eq:gbulk}) 
to determine ${\cal G}_0({\lambda})$ from the recurrence relation
${\cal G}_0({\lambda})^{-1}=\lambda - a' - b'^2{\cal G}({\lambda})$,
we obtain the following equation for the eigenvalue:
\begin{equation}
  \lambda_s - a' - b'^2 \left( 
  \frac{1}{2b^2} \left(\lambda_s - a \mp \sqrt{(\lambda_s -a)^2 - 4b^2}
  \right) \right) = 0 \: .
\label{eq:surfev}
\end{equation}
($\pm (\lambda_s-a) > 0$). 
Solving the equation for $\lambda_s$ yields the position of the 
eigenvalue in the spectrum of ${\bf K}$. Since ${\bf K}$ is real
and symmetric, only a solution $\lambda_s$ with 
$\mbox{Im} \, \lambda_s = 0$ is meaningful; a discrete 
$\lambda_s$ must lie
outside the bulk continuum (\ref{eq:cont}). Only the maximum 
eigenvalue in the spectrum is physically relevant 
[Eq.\ (\ref{eq:cond})]. Thus, we are interested in a solution 
that is split off the upper edge of the
continuum:
\begin{equation}
  \mbox{Re} \: \lambda_s > a + 2b \: .
\label{eq:split}
\end{equation}
Since $b\ge 0$ only the $-$ sign must be
considered in (\ref{eq:surfev}). 

Whether or not the condition (\ref{eq:split}) can be met, depends 
on the (surface) parameters $a'$ and $b'$. Solving Eq.\ 
(\ref{eq:surfev}) for $\lambda_s$ and inserting the solution into
(\ref{eq:split}), yields the following relation for $a'$ and $b'$
\begin{equation}
  2 b^2 + b (a-a') - b'^2 < 0 
  \: ,
\label{eq:sp}
\end{equation}
which must be fulfilled to obtain a (physically relevant) surface 
mode. Note that the relation cannot be satisfied with uniform 
parameters, i.~e.\ $a'=a$ and $b'=b$.

The interpretation is the following: In a semi-infinite system with 
surface parameters $t_{11}$, $t_{12}$ and $U_1$ that do not obey the 
condition (\ref{eq:sp}), there is only the ``ordinary'' transition
from a 
metallic to a Mott insulating state at $U=U_{c,{\rm bulk}}$ when 
increasing the interaction strength. The critical interaction 
$U_{c,{\rm bulk}}$ is given by Eq.\ (\ref{eq:ucbulk}). At this point 
all layer-dependent quasi-particle weights $z_\alpha$, in the bulk 
as well as 
at the surface, vanish. On the other hand, for a sufficiently 
strong modification of $t_{11}$, $t_{12}$ or $U_1$, i.~e.\ for $a'$ 
and $b'$ satisfying (\ref{eq:sp}), there are {\em two} critical 
interaction strengths: The first one is $U_{c,{\rm bulk}}$ again. 
At $U = U_{c,{\rm bulk}}$ the bulk quasi-particle weight 
$z_{\alpha = \infty}$ vanishes. The second critical interaction
strength $U_{c,{\rm surf}}$ can be determined from 
$\lambda_s \stackrel{!}{=} 1$ where $\lambda_s$ is the solution of 
(\ref{eq:surfev}). Let us assume that
$U_{c,{\rm surf}} > U_{c,{\rm bulk}}$. 
For $U > U_{c,{\rm surf}}$
the entire system is in the Mott insulating phase.
For $U_{c,{\rm bulk}} < U < U_{c,{\rm surf}}$, however, the bulk 
is a Mott insulator while the surface is still metallic.
We call the transition at $U=U_{c,{\rm bulk}}$ the ``extraordinary''
and the transition at $U=U_{c,{\rm surf}}$ the ``surface
transition'' in analogy with the terminology for 
magnetic phase transitions at surfaces \cite{Bin83}.

The remaining question is whether or not 
$U_{c,{\rm surf}} < U_{c,{\rm bulk}}$ can be possible. In such a 
situation we would have a quasi two-dimensional Mott insulator 
on top of a metallic bulk for interactions 
$U_{c,{\rm surf}} < U < U_{c,{\rm bulk}}$.
However, this possibility is ruled out: Eq.\ (\ref{eq:split}) can be 
rewritten as $\lambda_s > U^2_{c, {\rm bulk}} / U^2$. Furthermore,
at the critical point $U = U_{c, {\rm surf}}$ the value
$\lambda_s=1$ fulfills Eq.\ ({\ref{eq:surfev}). But 
this implies $1>U^2_{c,{\rm bulk}}/U^2_{c,{\rm surf}}$.
We can state that the linearized theory predicts that a metallic
surface coexisting with a Mott insulating bulk is possible while
the opposite scenario cannot be realized.

Arguing physically, if (at the Fermi edge) there is a finite (local) 
density of states in the second and all subsequent layers, this must 
always induce a non-zero, though possibly low density of states in 
the top layer, and thus an insulating surface phase is excluded: 
Consider the free-standing two-dimensional layer at an interaction 
strength $U_1$ being sufficiently strong to force the system to the 
insulating phase. Let the monolayer be coupled to the second and all 
subsequent layers by switching on the hopping between the top and 
the second layer $t_{12}\ne 0$. If $t_{12}$ is finite but too small, 
the low-energy bulk excitations cannot propagate into the top layer 
and are reflected at the Hubbard gap. However, virtual hopping 
processes are possible which cause (an exponentially damped) weight 
of bulk excitations in the top layer. The exponential damping 
becomes unimportant in this case since it is effective in one 
layer only. 

For the opposite case of a metallic surface on top of a Mott 
insulator, however, it does become essential: Low-energy excitations 
can propagate within the surface region since $U<U_{c, {\rm surf}}$. 
Because $U>U_{c, {\rm bulk}}$, they cannot propagate into the bulk
but are reflected at the (bulk) Hubbard gap. While virtual processes
always generate some non-zero spectral weight at the Fermi edge
in each layer, the weight is infinitesimally small asymptotically, 
for $\alpha \mapsto \infty$. 

Since critical fluctuations spread out all over the system at a 
second-order critical point, different parts of a system should 
undergo the transition at a common and unique critical value of the 
external control parameter. The exponential damping of low-energy 
excitations over large distances explains why there can be {\em two} 
critical interactions. This is analogous to the case of magnetic 
phase transitions at surfaces: In a system where a magnetic surface 
coexists with a paramagnetic bulk, the layer magnetization must
decay exponentially when passing from the surface to the crystal
volume. Vice versa: A magnetic bulk always induces a finite 
magnetization in the top layer. The exception is the somehow 
artificial case where the top layer is completely decoupled 
from the rest system (e.~g.\ $t_{12}=0$).

\subsection{Modified intra-layer surface hopping} 
\label{sec:mintra}

Some more aspects of the metallic surface phase shall be
addressed in the following. In particular, to discuss the effects 
of the surface geometry, we refer to the different low-index 
surfaces of the sc lattice mentioned above. 
Furthermore, it is helpful to consider the different types of 
surface  modifications separately.

We start by considering a modified intra-layer hopping in the top 
layer: $t_{11} \ne t$. We have:
\begin{equation}
  a' = \frac{t_{11}^2}{t^2} \, a \; , \;\; b' = b \: .
\end{equation}
From Eq.\ (\ref{eq:sp}) we can deduce that there are two critical
interactions provided that
\begin{equation}
  t_{11} > t \cdot \sqrt{1 + \frac{p}{q}} \: .
\label{eq:mm1}
\end{equation}
According to (\ref{eq:cond}) and (\ref{eq:surfev}), the critical 
interaction strength at which the surface transition takes place, 
is given by:
\begin{equation}
  U_{c,{\rm surf}} = 6t \cdot \sqrt{q \frac{t_{11}^2}{t^2} + 
  \frac{p^2}{q} \frac{t^2}{t_{11}^2 - t^2}} \: .
\label{eq:ucpara}
\end{equation}
The corresponding phase diagram is shown in Fig.~\ref{fig:para}. 

For the (111) surface there is no intra-layer hopping at all 
($q=0$). A rather moderate enhancement of $t_{11}$ (about 12\%) 
is sufficient to obtain a metallic 
surface phase for the sc(100) surface. In the case of the sc(110) 
surface a stronger modification is necessary. These trends are 
plausible: Obviously, for both surfaces a larger $t_{11}$ means 
that electrons in the top layer are more itinerant and thus tends 
to delay the transition to the insulating state as $U$ is increased. 
A smaller intra-layer coordination number $q$ counteracts this 
mechanism. Consequently, one needs a stronger enhancement of 
$t_{11}$ for the (110) surface. The $U$-range where a metallic 
surface coexists with an insulating bulk quickly increases at 
$t_{11}$ is increased. For $t_{11} \mapsto \infty$ one would 
expect that the energy scales relevant for the bulk become 
meaningless and that the electronic structure of the top layer
decouples from the rest system. This is predicted correctly by
Eq.\ (\ref{eq:ucpara}) which yields $U_{c,{\rm surf}} = 6t_{11} 
\sqrt{q}$ in this limit, i.~e.\ the critical interaction strength
of a free-standing two-dimensional layer.

\refstepcounter{figure}
\begin{figure}[t] 
\vspace{4mm}
\centerline{\psfig{figure=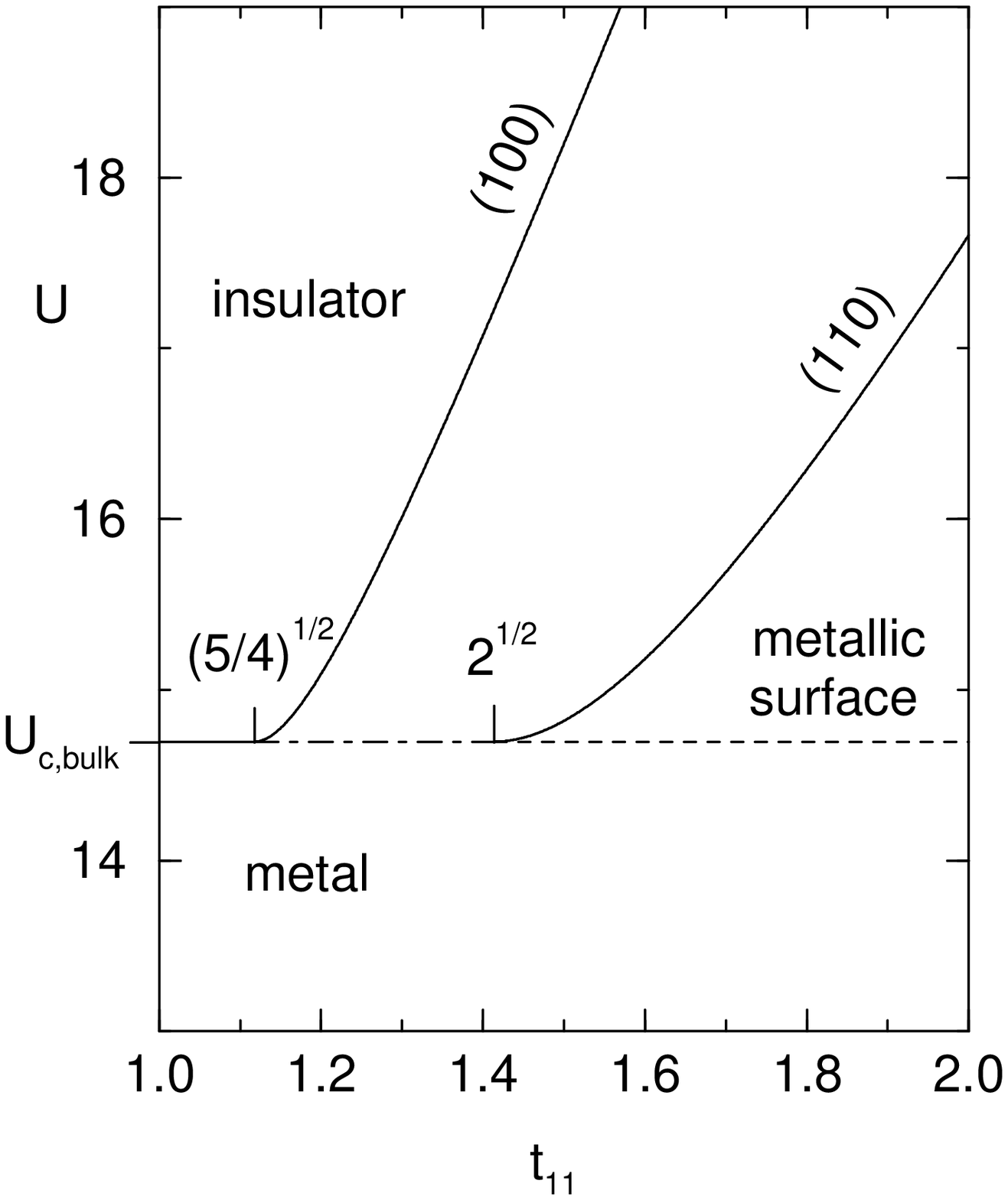,width=80mm,angle=0}}
\vspace{2mm}

\begin{center}
\parbox[]{120mm}{\small Fig.\ \ref{fig:para}:
$t_{11}$-$U$ phase diagram as obtained from the linearized 
DMFT. For $U<U_{c,{\rm bulk}}$ the system is metallic. For 
$U>U_{c,{\rm bulk}}$ the bulk is a Mott insulator, the surface
can be either insulating (left to the phase boundary)
or metallic (right). Phase boundaries for the (100) and (110)
surface of the sc lattice. Energy units: nearest-neighbor 
hopping $t=1$. Free band width $W=12$.
\label{fig:para}
}
\end{center}
\end{figure}

\subsection{Modified inter-layer surface hopping} 

For a modified inter-layer hopping between the top and the sub-surface
layer $t_{12} \ne t$ we have:
\begin{equation}
  a' = a \; , \;\; b' = \frac{t_{12}^2}{t^2} \, b \: .
\end{equation}
A metallic surface of a Mott insulating bulk is possible for
\begin{equation}
  t_{12} > \sqrt[4]{2} \, t \: ,
\label{eq:mm2}
\end{equation}
irrespective of the type of the surface.
The critical interaction strength for the surface transition 
is given by:
\begin{equation}
  U_{c,{\rm surf}} = 6t \cdot \sqrt{q + 
  p \frac{t_{12}^4}{t^2} \frac{1}{\sqrt{t_{12}^4-t^4}}} \: .
\label{eq:ucperp}
\end{equation}
Fig.~\ref{fig:perp} 
shows $U_{c,{\rm surf}}$ as a function of $t_{12}$ for the
different surfaces.

An enhancement of $t_{12}$ again means to enhance the itinerancy of 
electrons at the surface. Hopping processes between the topmost 
and the sub-surface layer become more likely. A modification of
about 19\% is sufficient to suppress the transition to the Mott
insulating phase at the surface for $U > U_{c,{\rm bulk}}$.
The surface critical interaction strength $U_{c,{\rm surf}}$ 
up to which the metallic surface phase persists for a given 
$t_{12}$, is the largest for the sc(111) surface since here 
the perpendicular hopping is favored by the comparatively 
high inter-layer coordination number $p=3$ anyway. In the limit 
$t_{12} \mapsto \infty$ the first two layers of the surface
will decouple from the bulk. The surface critical interaction
strength in this limit should be the same as for a bi-layer
system with strongly anisotropic hopping. Consider, for 
simplicity, the sc(111) surface where $q=0$. In this case
all sites in the bi-layer system have the same coordination
number $p$ and the bulk formula (\ref{eq:ucbulk}) may be applied
accordingly. This yields: $U_{c,{\rm surf}}=6t_{12}\sqrt{p}$, 
which is consistent with the $t_{12} \mapsto \infty$ limit 
of Eq.\ (\ref{eq:ucperp}).

\refstepcounter{figure}
\begin{figure}[t] 
\vspace{4mm}
\centerline{\psfig{figure=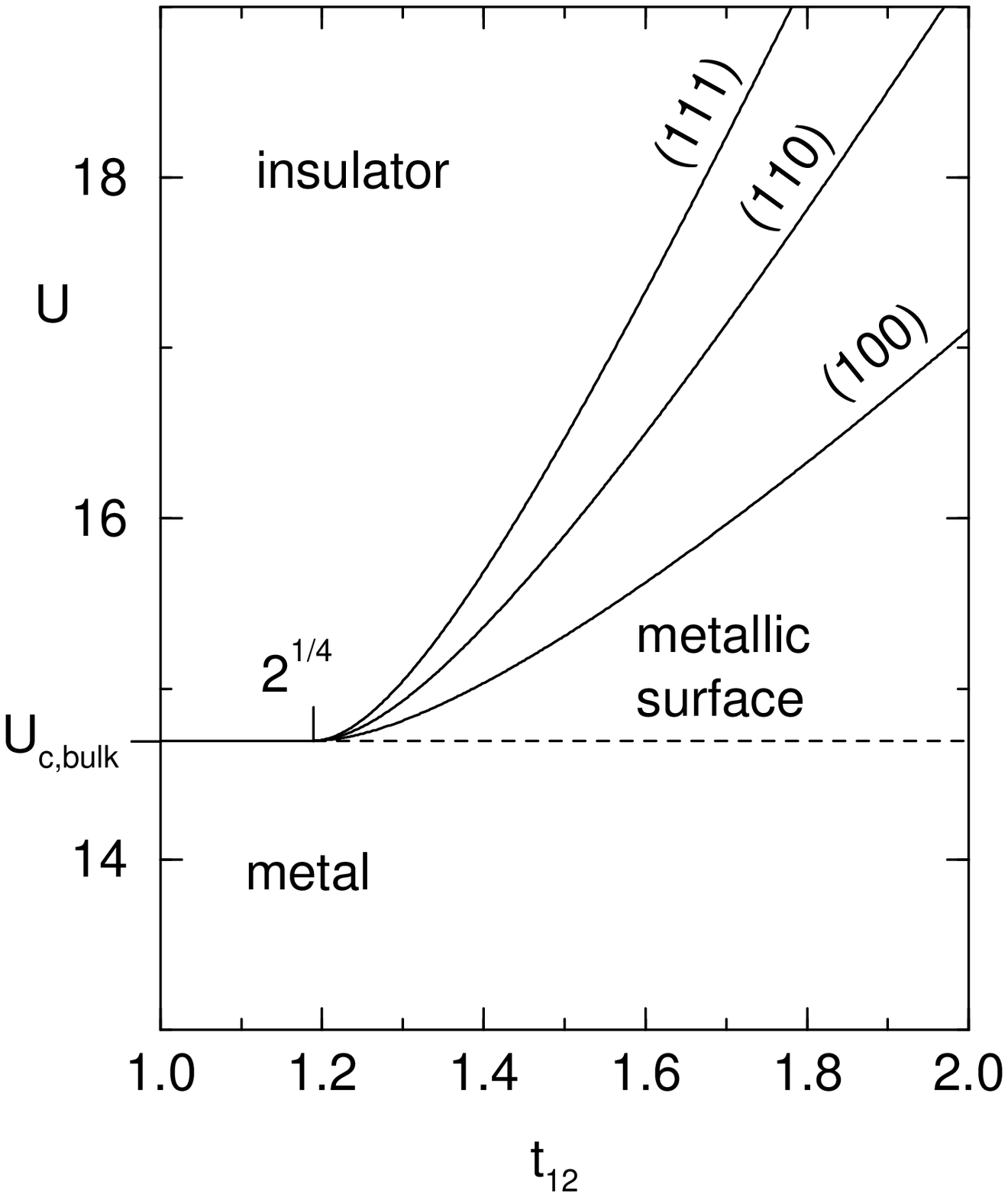,width=80mm,angle=0}}
\vspace{2mm}

\begin{center}
{\small Fig.\ \ref{fig:perp}:
$t_{12}$-$U$ phase diagram.
\label{fig:perp}
}
\end{center}
\end{figure}

\subsection{Modified surface Coulomb interaction} 

\refstepcounter{figure}
\begin{figure}[b] 
\vspace{4mm}
\centerline{\psfig{figure=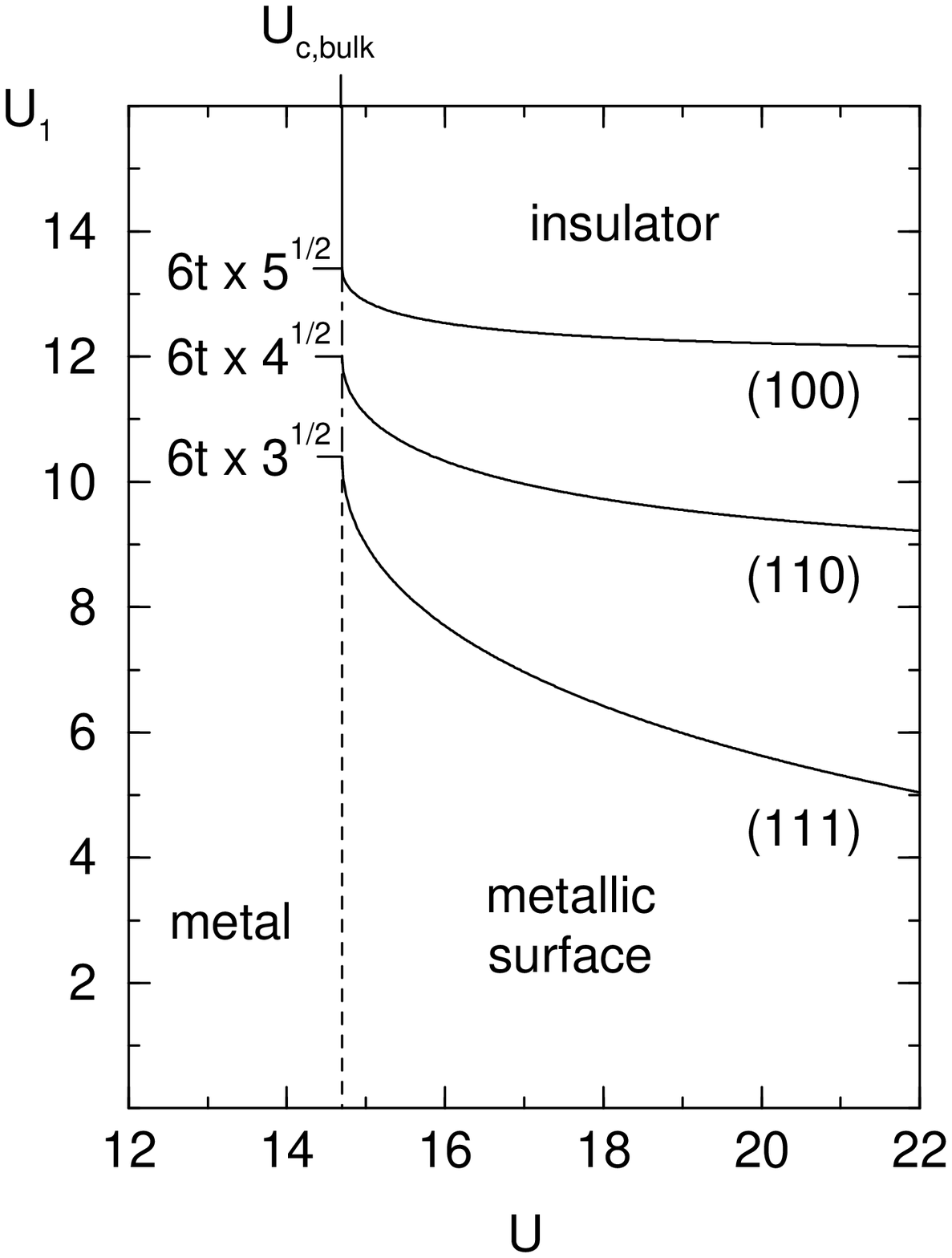,width=80mm,angle=0}}
\vspace{2mm}

\begin{center}
{\small Fig.\ \ref{fig:uuuu}:
$U$-$U_1$ phase diagram.
\label{fig:uuuu}
}
\end{center}
\end{figure}

Finally, we consider a modified Coulomb interaction in the top
layer, $U_1 \ne U$. In this case:
\begin{equation}
  a' = \frac{U^2}{U_1^2} \, a \; , \;\; b' = \frac{U}{U_1} b \: .
\end{equation}
As in the two other cases, we could fix the surface model
parameters, vary $U$ and ask for the critical interaction 
strength $U_{c,{\rm surf}}$. For the present case, however, it 
appears to be more intuitive to consider the bulk $U$ to be a 
fixed quantity and to vary $U_1$.

For $U$ above the bulk critical interaction $U_{c,{\rm bulk}}$
the bulk is a Mott insulator. The system may then become 
critical with respect to $U_1$, provided that
\begin{equation}
  U_1 < \sqrt{\frac{q+p}{q+2p}} \, U_{c, {\rm bulk}} \: .
\label{eq:ccuu}
\end{equation}
The surface transition takes place at $U_1=U_{1, c, {\rm surf}}$
with
\begin{equation}
  U_{1, c, {\rm surf}} = \sqrt{ 
  \frac{U^2 + 36 q t^2}{2} -
  \sqrt{\left( \frac{U^2 - 36 q t^2}{2} \right)^2 - 36^2 p^2 t^4}
  }
\label{eq:mm3}
\end{equation}
for $U>U_{c,{\rm bulk}}$. Fig.~\ref{fig:uuuu} 
shows the corresponding phase
diagram. For $U\mapsto \infty$ we get $U_{1, c, {\rm surf}} =
6t\sqrt{q}$. This is the critical interaction strength of the 
free-standing monolayer.

The results for modified surface Coulomb interaction can be compared
with Hasegawa's slave-boson approach \cite{Has92}. Qualitatively,
the respective $U$-$U_1$ phase diagrams for the sc(100) surface 
look similar. The critical interactions predicted by the slave-boson 
method are somewhat larger compared with the DMFT results. This is 
typical for the slave-boson method \cite{GKKR96}. An important
difference is found with respect to the ``special transition'' at
the tri-critical point 
$U=U_{c, {\rm bulk}}$, $U_1=U_{1,c} \equiv 
\sqrt{(q+p)/(q+2p)} ~ U_{c, {\rm bulk}}$.
The linearized DMFT predicts 
\begin{equation}
  \frac{U_{1, c, {\rm surf}} - U_{c,{\rm bulk}}}{U_{c,{\rm bulk}}}
  \propto \left(\frac{U_1}{U_{1,c}}-1\right)^{1/\phi}
\end{equation}
for $U_1 \mapsto U_{1,c}$ with a ``crossover exponent'' $\phi=1/2$. 
The same crossover exponent is found for modified surface hopping
$t_{11}$ or $t_{12}$:
\begin{equation}
  \frac{U_{c,{\rm surf}} - U_{c,{\rm bulk}}}{U_{c,{\rm bulk}}}
  \propto \left(\frac{t_{11(2)}}{t_{11(2),c}}-1\right)^{1/\phi} \: ,
\end{equation}
where $t_{11,c}$ and $t_{12,c}$ are defined by the r.h.s. of 
Eqs.\ (\ref{eq:mm1}) and (\ref{eq:mm2}), respectively.
This follows from a 
direct calculation and can also be seen in Figs.\ \ref{fig:para},
\ref{fig:perp} and \ref{fig:uuuu}.
Contrary, within the slave-boson theory of Ref.\ \cite{Has92}, 
$U_{1, c, {\rm surf}}$ seems to be 
constant as a function of $U$, and a
crossover exponent cannot be defined.

\subsection{Profiles of the quasi-particle weight} 

The mean-field equation of the linearized DMFT, 
$\Delta_{N+1}^{(\alpha)} = \sum_\beta K_{\alpha \beta} 
\Delta_N^{(\beta)}$,
has a non-trivial solution only at a critical point for the
Mott transition, e.~g.\ at $U=U_{c,{\rm bulk}}$ or 
$U=U_{c,{\rm surf}}$. This solution is a fixed point of the matrix
$\bf K$, $\Delta_\infty^{(\alpha)} = \lim_{N\mapsto \infty}
\Delta_N^{(\alpha)}$, and can be calculated as the eigenvector 
of ${\bf K}$ belonging to the eigenvalue $\lambda=1$ 
[Eq.\ (\ref{eq:cond})]. 
Since $z_\alpha \propto \Delta^{(\alpha)}$,
the eigenvector has the meaning
of the critical profile of the quasi-particle weight, i.~e.\ the
$\alpha$-dependence of $z_\alpha$ in the limit $z_\alpha\mapsto 0$.
It is uniquely determined up to a normalization constant.

The upper left part of Fig.\ \ref{fig:prof} shows the critical
profile at the sc(100) surface for different values of $t_{11}$
and $U=U_{c,{\rm bulk}}$ or $U=U_{c,{\rm surf}}$, respectively.
$z_\alpha$ has been normalized to its top-layer value $z_1$.
For unmodified surface hopping $t_{11}=t$, the profile is linear.
In fact, the ansatz $z_\alpha \propto \alpha$ solves the mean-field 
equation $z_\alpha=(36t^2/U^2)(qz_\alpha+pz_{\alpha+1}+pz_{\alpha-1})$ 
for $U=U_{c, {\rm bulk}}=6t\sqrt{q+2p}$. Physically, this means that
at the critical interaction the surface effects extend into the bulk 
up to {\em arbitrarily} large distances. Note that this implies that 
actually an infinite number of inequivalent surface layers has to be
considered in a fully numerical evaluation of the DMFT.

\refstepcounter{figure}
\begin{figure}[t] 
\vspace{4mm}
\centerline{\psfig{figure=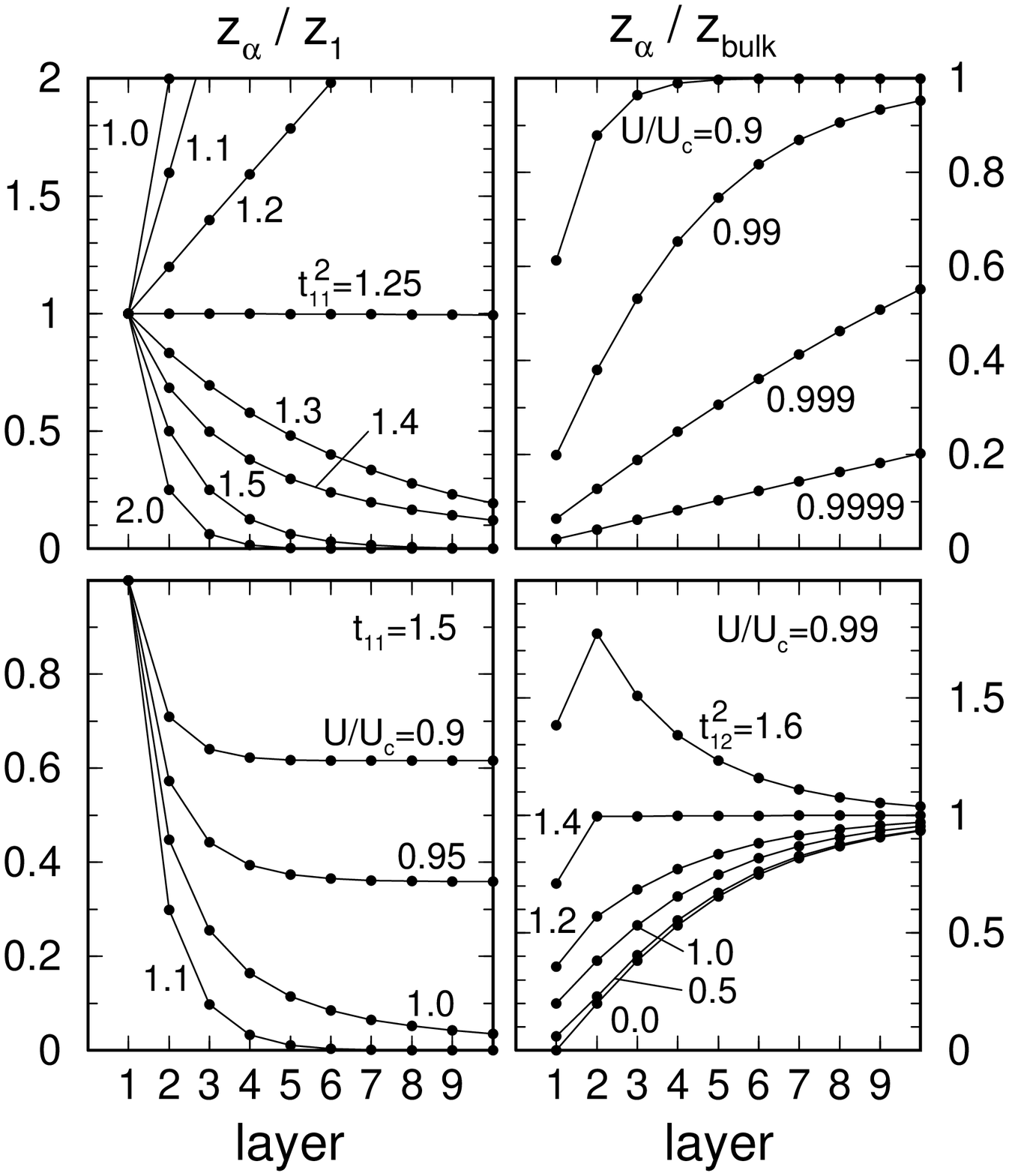,width=100mm,angle=0}}
\vspace{2mm}

\begin{center}
\parbox[]{150mm}{\small Fig.\ \ref{fig:prof}:
Profiles of the quasi-particle weight for the sc(100) surface.
{\em Upper left:} Profiles for different $t_{11}$. 
$U=U_{c ,{\rm bulk}}$ for $t_{11} \le 1.25$ and
$U=U_{c ,{\rm surf}}$ for $t_{11} \ge 1.25$. 
{\em Upper right:} $t_{11}=t=1$ and different $U$ close to 
$U_{c ,{\rm bulk}}$.
{\em Lower left:} 
$t_{11}=1.5$ and different $U/U_{c ,{\rm bulk}}$.
{\em Lower right:} 
$U/U_{c ,{\rm bulk}}=0.99$ and different $t_{12}^2$.
The profiles are normalized to the top-layer
value (r.h.s.) or the bulk value (l.h.s.), respectively.
\label{fig:prof}
}
\end{center}
\end{figure}

For $U$ close to $U_{c, {\rm bulk}}$ but $U < U_{c, {\rm bulk}}$, 
one would expect that the profile converges to a finite bulk value: 
$\lim_{\alpha \mapsto \infty} z_\alpha = z_{\rm bulk} > 0$. In its 
present form, however, the linearized DMFT is not applicable here. 
One may consider the following extension \cite{BP99} of the mean-field 
equation (for simplicity, we discuss the case of uniform model 
parameters, the generalization for modified surface hopping or 
$U_1 \ne U$ is straightforward):
\begin{equation}
  z_\alpha=\frac{36t^2}{U^2} 
  \left( qz_\alpha+pz_{\alpha+1}+pz_{\alpha-1} \right)
  - c~z_\alpha^2
  \: .
\label{eq:landau}
\end{equation}
A quadratic term in $z_\alpha$ with a constant coefficient $c>0$ has 
been added. The constant $c$ can be fixed by the value for 
$z_{\rm bulk}$ or for $z_1$ (Ref.\ \cite{BP99} yields the explicit
value $c=11/9$ but we do not need the result here). 
This extension of the linearized
DMFT is in the spirit of Landau theory, we simply consider the 
next term in an expansion with respect to the ``order parameter'' 
$z_\alpha$. As in the Landau theory, higher-order terms in $z_\alpha$ 
or quadratic terms that couple the different layers are still 
neglected. The additional term in Eq.\ (\ref{eq:landau}) ensures 
a linear $U$-dependence of the quasi-particle weight in each layer: 
$z_\alpha \propto (U_c - U)$ for $U\mapsto U_c$. This is consistent 
with the (bulk) critical behavior found within the PSCM \cite{MSK+95}.

Using Eq.\ (\ref{eq:landau}) we have calculated the profile of the
quasi-particle weight for $t_{11}=1$ and different 
$U<U_{c,{\rm bulk}}$, see Fig.\ \ref{fig:prof} (upper right).
For $U/U_{c,{\rm bulk}} = 0.9$ the quasi-particle weight 
significantly differs from the bulk value in the first few
layers from the surface only. As $U\mapsto U_{c,{\rm bulk}}$, 
however, the linear trend of $z_\alpha$ clearly develops.

A linear trend of the critical profile is also observed for slightly 
enhanced surface hopping, $t^2_{11} = 1.1$ and $t^2_{11} = 1.2$ 
(Fig.\ \ref{fig:prof}, upper left).
For a surface hopping $t_{11} = \sqrt{1+p/q} = \sqrt{5/4}$ 
we get the so-called special transition (cf.\ Eq.\ (\ref{eq:mm1}) 
and Fig.\ \ref{fig:para}). At the critical interaction the 
profile is a constant (Fig.\ \ref{fig:prof}, upper left). 
In this case the effect of missing neighbors at the surface is 
exactly compensated by the enhancement of $t_{11}$.

For $t_{11} > \sqrt{5/4}$ there are two critical interactions, 
$U_{c,{\rm bulk}}$ and $U_{c,{\rm surf}}$. For $U=U_{c,{\rm surf}}$, 
$z_\alpha/z_1$ is at its maximum in the top layer and exponentially 
decays as $\alpha \mapsto \infty$ (Fig.\ \ref{fig:prof}, upper left). 
For $U < U_{c,{\rm surf}}$ [according
to Eq.\ (\ref{eq:landau})] the decay becomes slower until the 
profile converges to a finite bulk value for $U<U_{c,{\rm bulk}}$
(lower left).

Finally, the lower right part of Fig.\ \ref{fig:prof} shows the 
profile of the quasi-particle weight obtained from Eq.\ 
(\ref{eq:landau}) for $U/U_{c,{\rm bulk}} = 0.99$ and modified
inter-layer surface hopping $t_{12}$. For $t_{12}^2 < \sqrt{2}$
the profile is a monotonously increasing function when passing from
the surface to the bulk. $t_{12}^2 = \sqrt{2}$ marks the special
transition [see Eq.\ (\ref{eq:mm2})]. Here the profile would be
constant for $\alpha \ge 2$ and $U=U_{c, {\rm bulk}}$ as can be
seen from the mean-field equation of the linearized DMFT. For 
$t_{12}^2 > \sqrt{2}$ the quasi-particle weight is enhanced at 
the surface and monotonously decreases for $\alpha \ge 2$.

\subsection{Infinite dimensions} 

Dynamical mean-field theory rests on the local approximation for
the self-energy functional. 
Since it is known that in the limit of high spatial 
dimensions $D\mapsto \infty$ \cite{MV89} the local approximation 
becomes exact \cite{MH89b}, it may be interesting to discuss the 
(somewhat artificial) case of a surface of the infinite-dimensional 
hyper-cubic lattice. 

A $D$-dimensional hyper-cubic lattice may be thought to be built
up from $(D-1)$-dimensional ``layers'' perpendicular to a 
$D$-dimensional spatial direction characterized by the set of 
Miller indices $[x_1,x_2, \dots ,x_D]$. Cutting the hopping 
between two adjacent layers, one obtains a 
``$(x_1,x_2, \dots ,x_D)$ surface''. Consider the low-index 
directions with
$x_1 = \cdots = x_r = 1$ and $x_{r+1} = \cdots = x_D = 0$. 
For a given site there are $q=2D-2r$ nearest neighbors within 
the same layer and $p=r$ nearest neighbors in each of the adjacent
layers ($Z=q+2p=2D$).

For $r=1$, i.~e.\ a (1000...) surface, a site in the topmost 
layer has $Z_{\rm S}=q+p=2D-1$ nearest neighbors to be compared 
with $Z=2D$ in the bulk. For $D \mapsto \infty$ the local environment 
of the surface sites is essentially the same as in the bulk, 
surface effects become meaningless. With the usual scaling of
the hopping $t = t^\ast / \sqrt{2D}$ \cite{MV89}, the free top-layer 
local density of states (DOS) is a Gaussian 
$\rho^{(0)}(E)=\exp[-(E/t^\ast)^2/2]/(\sqrt{2\pi} t^\ast)$ --
as in the bulk. 

For $r=D$ one obtains the open (1111...) surface.
The surface coordination number is reduced to $Z_{\rm S}=p=D$.
This implies a ratio 
$\Delta_{\rm S}/\Delta = Z_{\rm S}/Z = (q+p)/(q+2p) = 0.5$ between 
the variances of the top-layer and bulk DOS. The results of a simple 
numerical calculation are shown in Fig.~\ref{fig:inf1111}. 
We notice a strongly 
modified and strongly layer-dependent DOS near the surface which 
slowly converges to the bulk Gaussian DOS for $\alpha \mapsto \infty$. 
In many respects the results resemble the DOS at the $D=3$ sc(111) 
surface, in particular the oscillation of $\rho^{(0)}_\alpha(E=0)$ 
as a function of $\alpha$ \cite{KS71}.

In infinite dimensions dynamical mean-field theory is exact also 
for the semi-infinite model. The scaling of the hopping implies
$G^{(0)}_{ij} \sim 1/\sqrt{D}$ for the free propagator between 
arbitrary nearest-neighbor sites $i$ and $j$, and the proof that 
the self-energy is local, is essentially unchanged (see Refs.\ 
\cite{MV89,Vol93,MH89b}). The simple linearized DMFT can be 
developed as in Sec.\ \ref{sec:s4}. 
We only have to insert the general 
expressions for the coordination numbers $q=2D-2r$ and $p=D$, 
and to perform the limit $D\mapsto \infty$ in the equations 
(\ref{eq:mm1}) to (\ref{eq:mm3}), paying attention to the scaling 
of the hopping.

\refstepcounter{figure}
\begin{figure}[t] 
\vspace{4mm}
\centerline{\psfig{figure=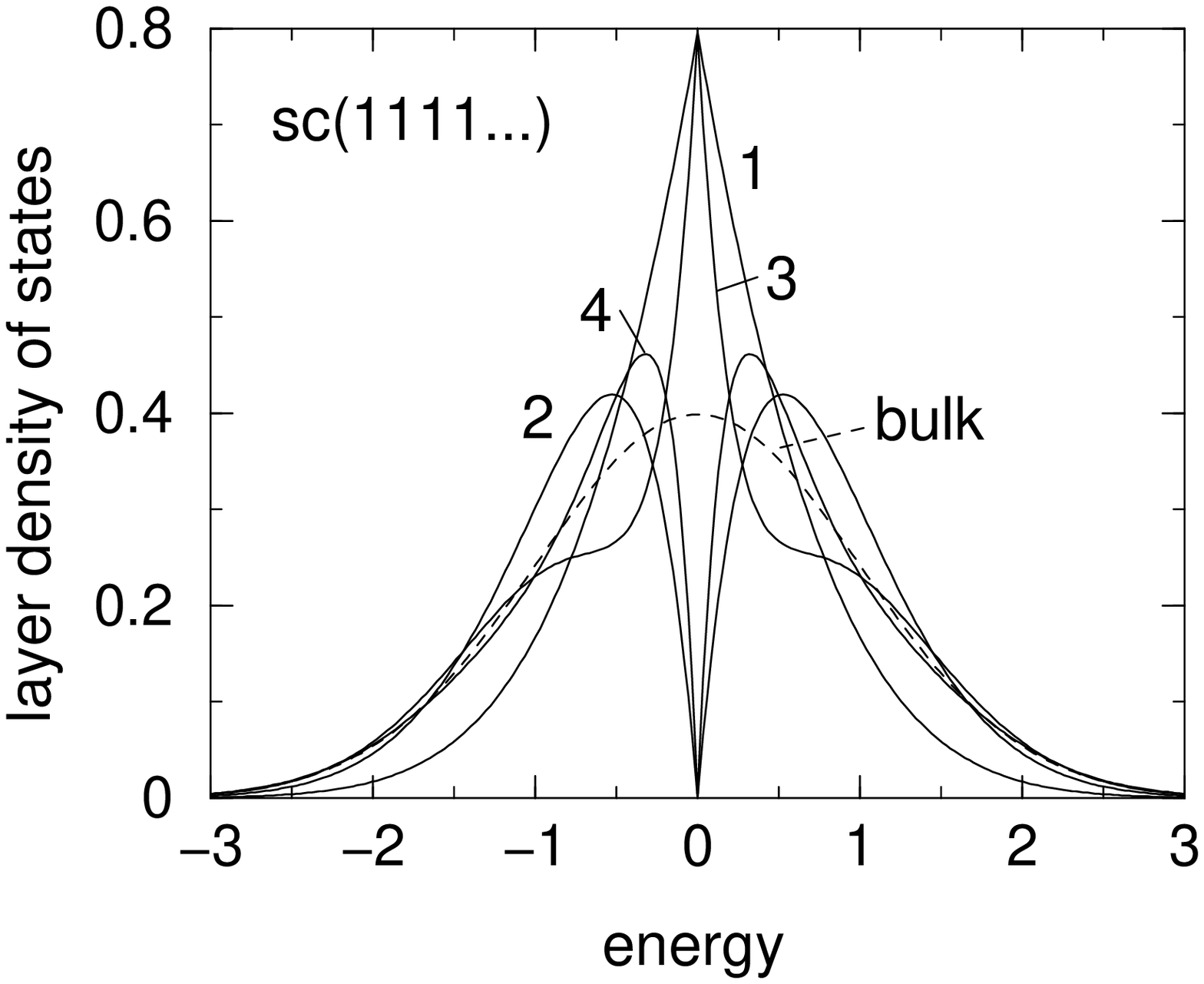,width=80mm,angle=0}}
\vspace{2mm}

\begin{center}
\parbox[]{120mm}{\small Fig.\ \ref{fig:inf1111}:
$U=0$ layer-dependent density of states 
$\rho^{(0)}_\alpha(E)$ at the $(1111...)$ surface
of the $D=\infty$ hyper-cubic lattice. Scaled hopping 
$t=t^\ast/\sqrt{2D}$ with $t^\ast=1$. ``1'' stands for the
topmost surface layer, ``2'' denotes the sub-surface layer, etc.
\label{fig:inf1111}
}
\end{center}
\end{figure}

Varying $r$ we can then pass continuously from the most closed 
($r=1$) to the most open ($r=D$) surface geometry. Consider, for
example, a modified intra-layer surface hopping. A surface phase 
is predicted to be existing for $t_{11}^\ast > t^\ast 
\sqrt{1+r/(2D-2r)}$ [cf.\ Eq.\ (\ref{eq:mm1})]; i.~e.\ 
for all $t_{11}^\ast > t^\ast$ in the case of the closed $r=1$
surface and not at all for the $r=D$ surface. For $r=1$ the
surface critical interaction is given by 
$U_{c,{\rm surf}} = 6 t_{11}^\ast$ [Eq.\ (\ref{eq:ucpara})] to be 
compared with the bulk critical interaction 
$U_{c,{\rm bulk}} = 6 t^\ast$ [Eq.\ (\ref{eq:ucbulk})]. 
With increasing $r$, $U_{c,{\rm surf}}$ decreases until
$U_{c,{\rm surf}} = U_{c,{\rm bulk}}$ for $r=D$.

The other cases may be discussed accordingly. Upon taking the
limit $D\mapsto \infty$, we always obtain non-trivial and plausible 
results. The discussion is analogous to the $D=3$ case. We conclude 
that the semi-infinite Hubbard model remains non-trivial for 
$D=\infty$ and provides a useful framework for investigating the 
surface phase. In principle, this can be done without approximations 
by employing the DMFT. Recall, however, that the linearized DMFT is 
still approximate (Sec.\ \ref{sec:s4} and Ref.\ \cite{BP99}).

\section{Exact-diagonalization method} 
\label{sec:s6}

For a complete numerical solution of the mean-field equations at 
finite temperatures one may employ the Quantum-Monte-Carlo method 
\cite{Jar92,RZK92,GK92b}. For $T=0$ the Exact-Diagonalization (ED) 
approach \cite{CK94,SRKR94,RMK94} can be applied and is chosen 
here. The main idea is to map onto a SIAM with a finite number 
of sites $n_s$. Lancz\`os technique \cite{Hay80} is used to 
calculate the ground state as well as the $T=0$ impurity Green 
function and self-energy. The DMFT equations are solved on the
discrete mesh of Matsubara energies where the (fictitious) inverse
temperature $\widetilde{\beta}$ introduces a low-energy cutoff.
Details of the method can be found in Ref.\ \cite{GKKR96}. The 
surface geometry can be simulated by a slab consisting of a finite 
but sufficiently large number of layers $d$ (for $U\ne U_c$). 
The numerical effort then increases linearly with $d$ at least. 
In Refs.\ \cite{PN99a,PN99c} we have discussed the application of 
ED to film and surface geometries.

ED is able to yield the essentially exact solution of the mean-field 
equations in a parameter range where the errors introduced by the 
finite system size are unimportant. For the Mott problem the relevant 
low-energy scale is set by the width of the quasi-particle peak 
in the metallic solution. It has to be expected that there are 
non-negligible finite-size effects when this energy scale becomes 
comparatively small. We are thus limited to interactions strengths 
that are not too close to $U_{c,{\rm bulk}}$ or $U_{c,{\rm surf}}$ 
and cannot access the very critical regimes. This also implies that
a precise determination of $U_{c,{\rm bulk}}$ and $U_{c,{\rm surf}}$ 
and thereby a direct comparison with the linearized DMFT is not 
possible. The discussion in \cite{PN99a}, however, shows that the
main trends can be derived safely.

In the following we mainly focus on the low-energy electronic 
structure which the ED method is able to predict reliably in the 
non-critical regimes. The so-called layer-dependent quasi-particle 
weight,
\begin{equation}
  z_\alpha = \left( 1 - \frac{d \Sigma_{\alpha}(E=0)}{dE} 
  \right)^{-1} \: ,
\end{equation}
is the primary quantity of interest. $z_\alpha \le 1$ is weight of 
the coherent quasi-particle peak in the local DOS 
$\rho_\alpha(E)$ of the $\alpha$-th layer or, alternatively, the
reduction factor of the discontinuous drops in $\alpha$-th 
momentum-distribution function $n_\alpha({\bf k})$ when $\bf k$ 
crosses the one-dimensional Fermi ``surfaces'' \cite{PN99a}.

Routinely, the calculations have been performed with $n_s=8$ sites 
in the effective impurity problems. For the fictitious temperature 
we have chosen $\widetilde{\beta}^{-1} = 0.0016 \, W$ ($W=12$ is
the free band width). $n_s$ and $\widetilde{\beta}$ determine the
``energy resolution'' which is found to be about 
$\Delta E = 0.12 = W/100$.
This implies that reliable results can be expected in a parameter 
region where $z_\alpha > 0.01$ (cf.\ Ref.\ \cite{PN99c}).
A moderate number $d\le 25$ of layers in the slab is sufficient 
to simulate the semi-infinite system -- except for the very critical
regime. This has been checked by 
comparing the results from calculations for different $d$.
We made use of the mirror symmetry at the center of the slab 
and of electron-hole symmetry to reduce the number of parameters,
the conduction-band energies $\epsilon_k$ and the hybridization 
strengths $V_k$ ($k=2,...,n_s$), which have to be determined 
self-consistently. We always found a unique and fully stabilized 
solution.

\section{Numerical results for the sc(100) surface} 
\label{sec:s7}

To keep the calculations manageable, we restrict the discussion to 
the $D=3$ sc(100) surface in the following. We start with the case 
of uniform model parameters. Fig.~\ref{fig:zofu} 
shows the bulk quasi-particle
weight $z$ (dashed line) as a function of $U$. It starts from its 
non-interacting value $z=1$. A quadratic $U$ dependence is noticed 
for small $U$ in agreement with perturbation theory \cite{PN99c}.
$z$ vanishes as $U$ approaches $U_{c,{\rm bulk}}$. The overall
dependence on $U$ is very similar to what is known from DMFT studies
of the $D=\infty$ Bethe lattice \cite{GKKR96}. 

In the top layer of the sc(100) surface the quasi-particle weight 
is significantly reduced (solid line). The lowered coordination 
number at the surface implies a reduced variance $\Delta_{\rm S}$ 
of the free surface DOS and thereby an increased 
effective interaction $U/\sqrt{\Delta_{\rm S}}$ compared with the bulk. 
Thus, at the surface correlation effects are enhanced, and 
$z_{\alpha=1}$ is lowered. Despite this {\em tendency} towards 
an insulating surface, we find a common critical interaction 
$U_{c,{\rm surf}}=U_{c,{\rm bulk}}$ which, for uniform parameters,
is in agreement with the analytical results. $U_{c,{\rm bulk}}$
also represents the critical interaction for all sub-surface layers.
For the rather closed (100) surface, $z_\alpha(U)$ is almost 
identical with the bulk function for $\alpha \ge 2$. 

\refstepcounter{figure}
\begin{figure}[t] 
\vspace{4mm}
\centerline{\psfig{figure=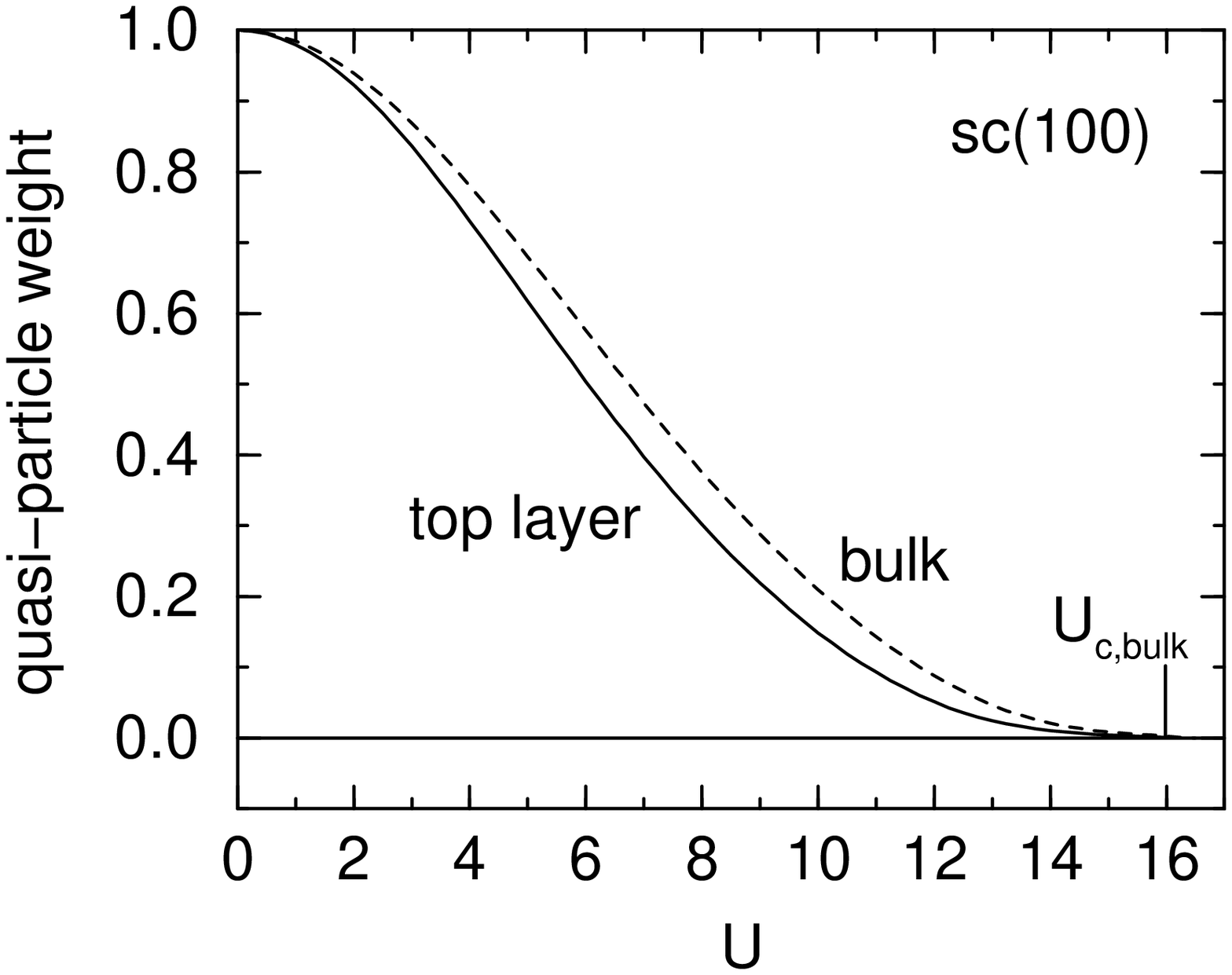,width=80mm,angle=0}}
\vspace{2mm}

\begin{center}
\parbox[]{120mm}{\small Fig.\ \ref{fig:zofu}:
$U$ dependence of the quasi-particle weight in the bulk and in the
top layer for the sc(100) surface (uniform model parameters) as 
obtained from the ED method for $n_s=8$. 
$U_{c, {\rm bulk}} \approx 16.0$. $t=1$ sets the energy scale.
}
\end{center}
\label{fig:zofu}
\end{figure}

From Fig.~\ref{fig:zofu} 
we can read off $U_{c,{\rm bulk}} \approx 16.0$ while
Eq.\ (\ref{eq:ucbulk}) predicts $U_{c,{\rm bulk}} = 14.7$. We have
to bear in mind, however, the underlying assumptions that lead to 
(\ref{eq:ucbulk}). Moreover, as concerns the ED, finite-size 
effects prevent a precise estimate: $U_{c,{\rm bulk}} \approx 15.1$ 
is found for $n_s=10$ sites in the impurity models \cite{PN99c}. 
On the other hand, comparing the results for $n_s=8$ and $n_s=10$, 
there are no significant changes as long as $z_\alpha > 0.01$
\cite{PN99c}. This means (see Fig.~\ref{fig:zofu}) 
that the overall layer and 
$U$ dependence is predicted reliably. We also believe that the
finding of a common critical interaction is not an artifact of
the ED approach since this is made plausible by the linearized
DMFT.

At the critical interaction the metallic solution continuously 
coalesces with the insulating solution that is found for 
$U > U_{c, {\rm bulk}}$. The insulating solution persists
down to another (common) critical interaction strength 
$U_{c,1} < U_{c, {\rm bulk}}$ (we find $U_{c,1} \approx 11.5$). 
In the coexistence region, however, it is thermodynamically 
irrelevant. For details we refer to Refs.\ 
\cite{GKKR96,RMK94,Bul99,PN99a}.

\subsection{Modified intra-layer surface hopping} 

\refstepcounter{figure}
\begin{figure}[t] 
\vspace{4mm}
\centerline{\psfig{figure=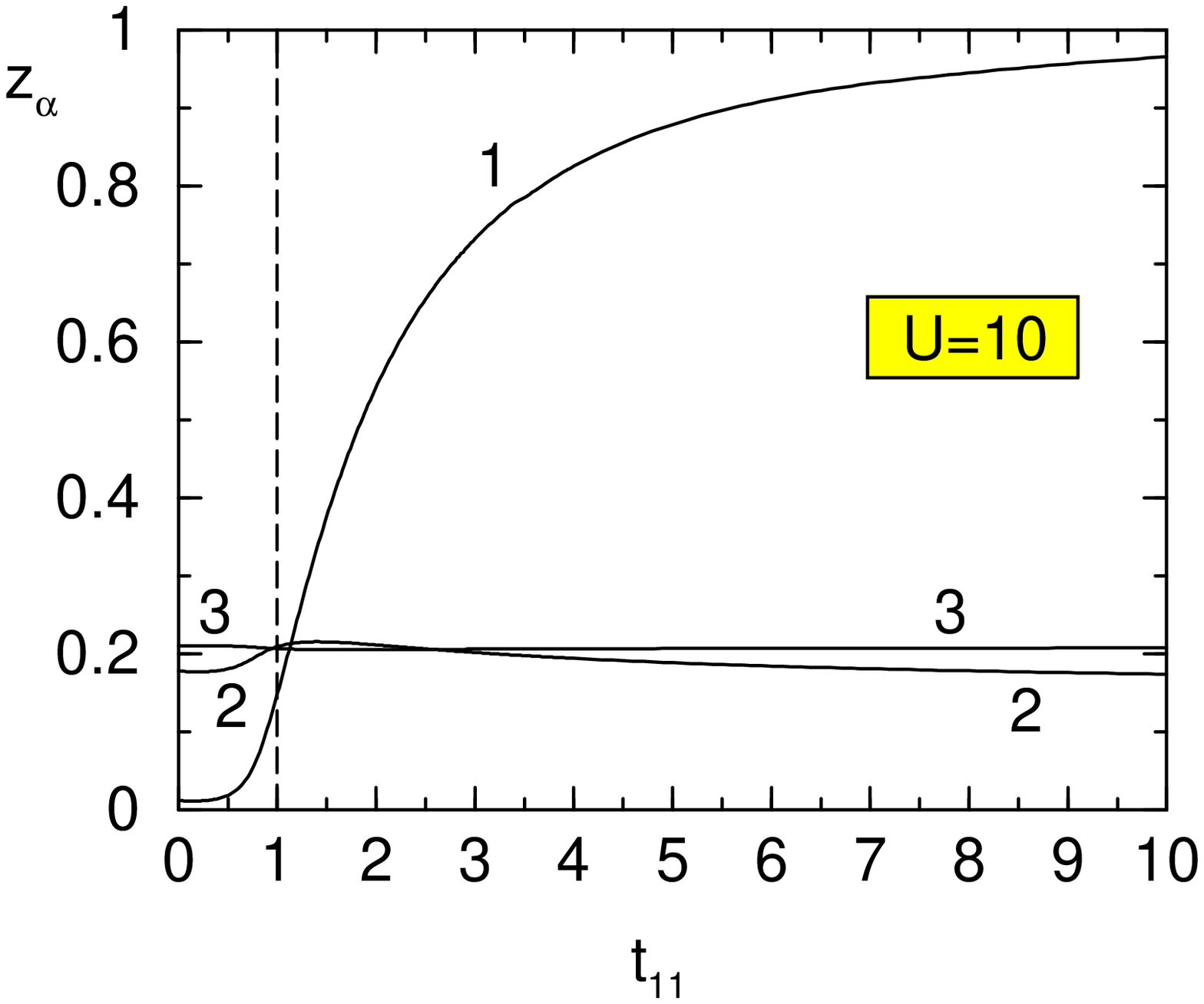,width=80mm,angle=0}}
\vspace{2mm}

\begin{center}
\parbox[]{120mm}{\small Fig.\ \ref{fig:para1}:
Quasi-particle weight of the top layer ($\alpha=1$) and the 
sub-surface layers ($\alpha=2,3$) for $U=10<U_{c, {\rm bulk}}$ 
as a function of the modified intra-layer surface hopping $t_{11}$. 
$t=1$.
}
\end{center}
\label{fig:para1}
\end{figure}

\refstepcounter{figure}
\begin{figure}[ht] 
\vspace{4mm}
\centerline{\psfig{figure=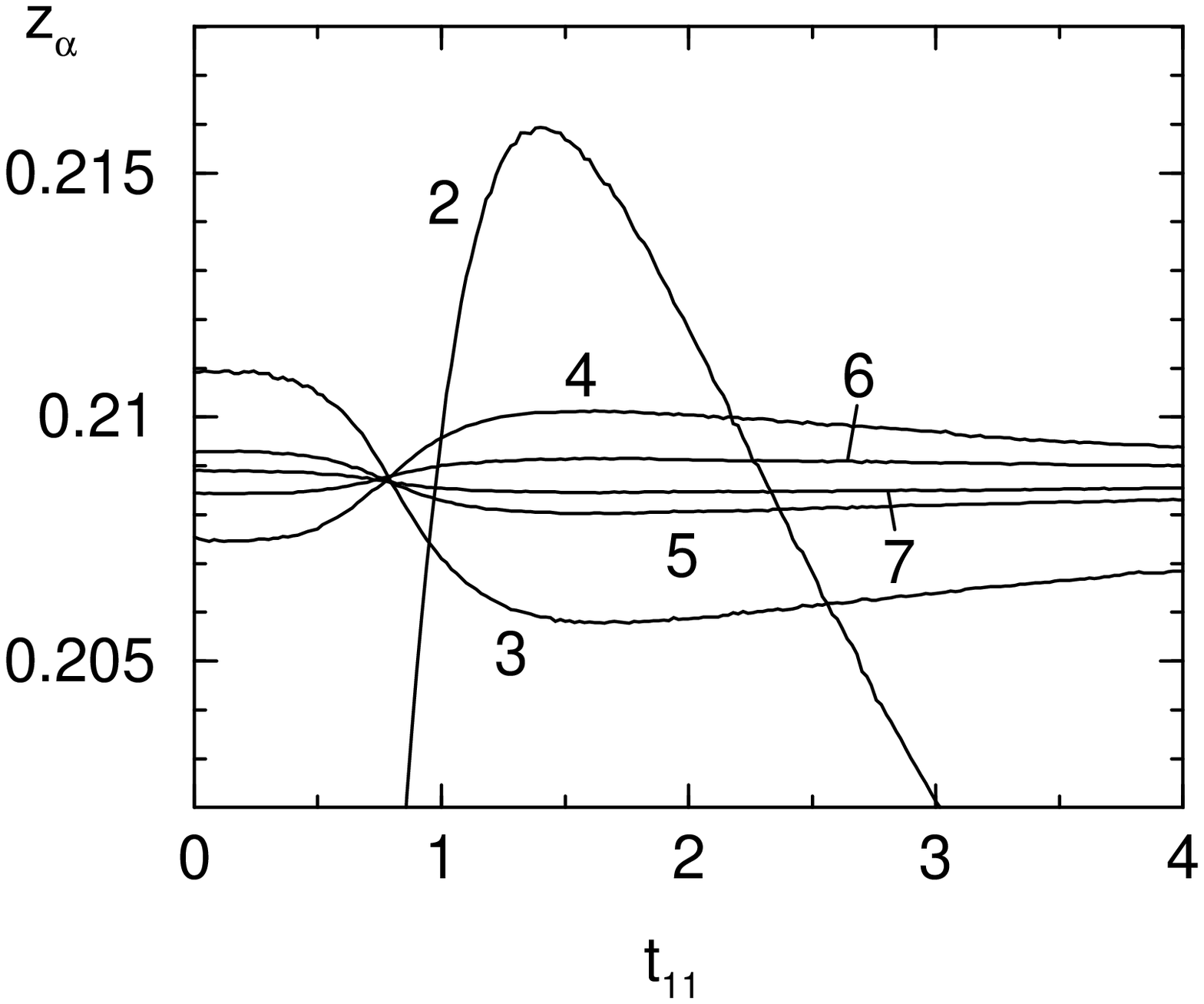,width=80mm,angle=0}}
\vspace{2mm}

\begin{center}
\parbox[]{120mm}{\small Fig.\ \ref{fig:para2}:
The same as Fig.~\ref{fig:para1} but on an enlarged scale.
}
\end{center}
\label{fig:para2}
\end{figure}

\refstepcounter{figure}
\begin{figure}[t] 
\vspace{4mm}
\centerline{\psfig{figure=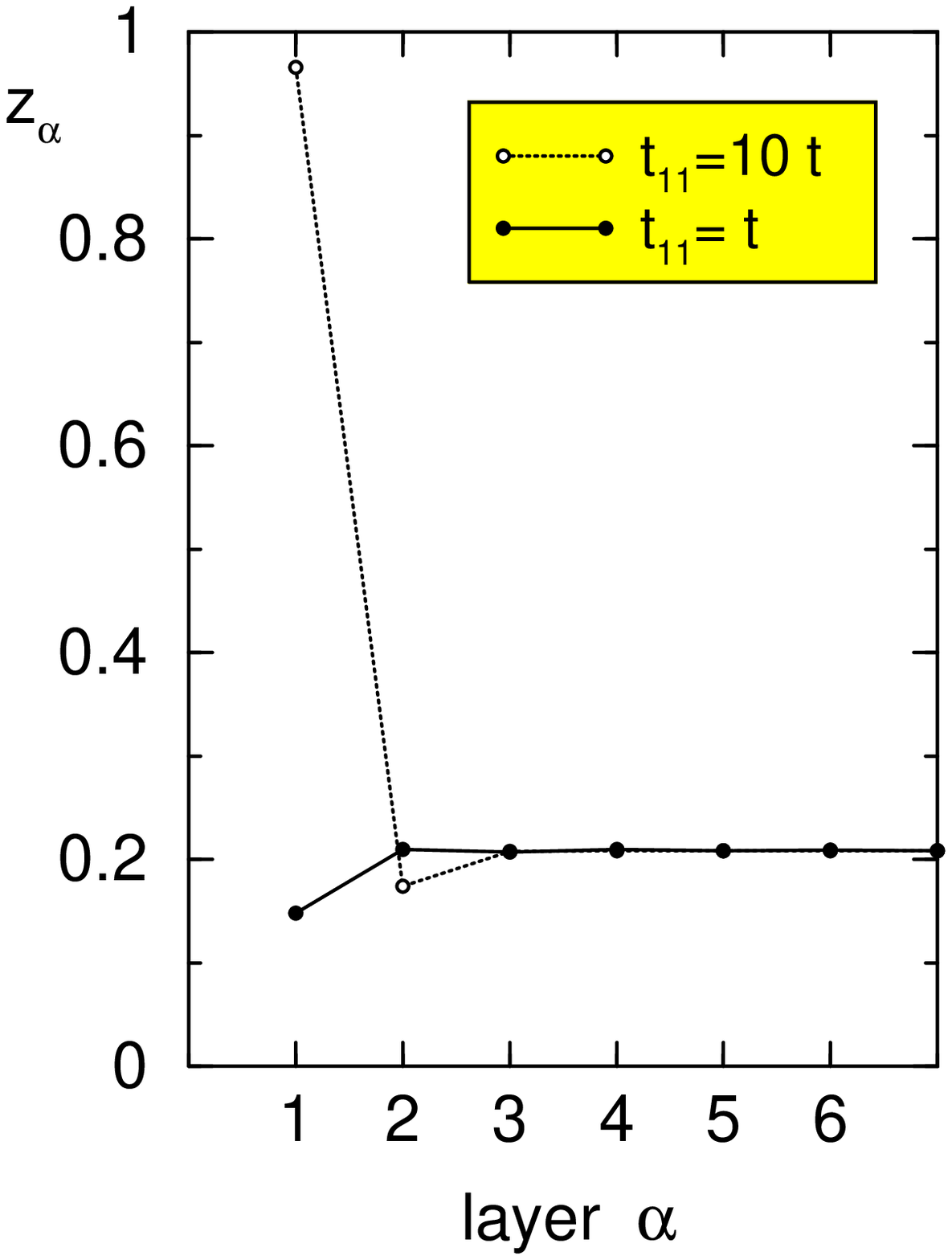,width=70mm,angle=0}}
\vspace{2mm}

\begin{center}
\parbox[]{105mm}{\small Fig.\ \ref{fig:para4}:
Layer-dependence of the quasi-particle weight for uniform model
parameters as well as strongly enhanced intra-layer surface 
hopping. $U=10$.
}
\end{center}
\label{fig:para4}
\end{figure}

\refstepcounter{figure}
\begin{figure}[t] 
\vspace{4mm}
\centerline{\psfig{figure=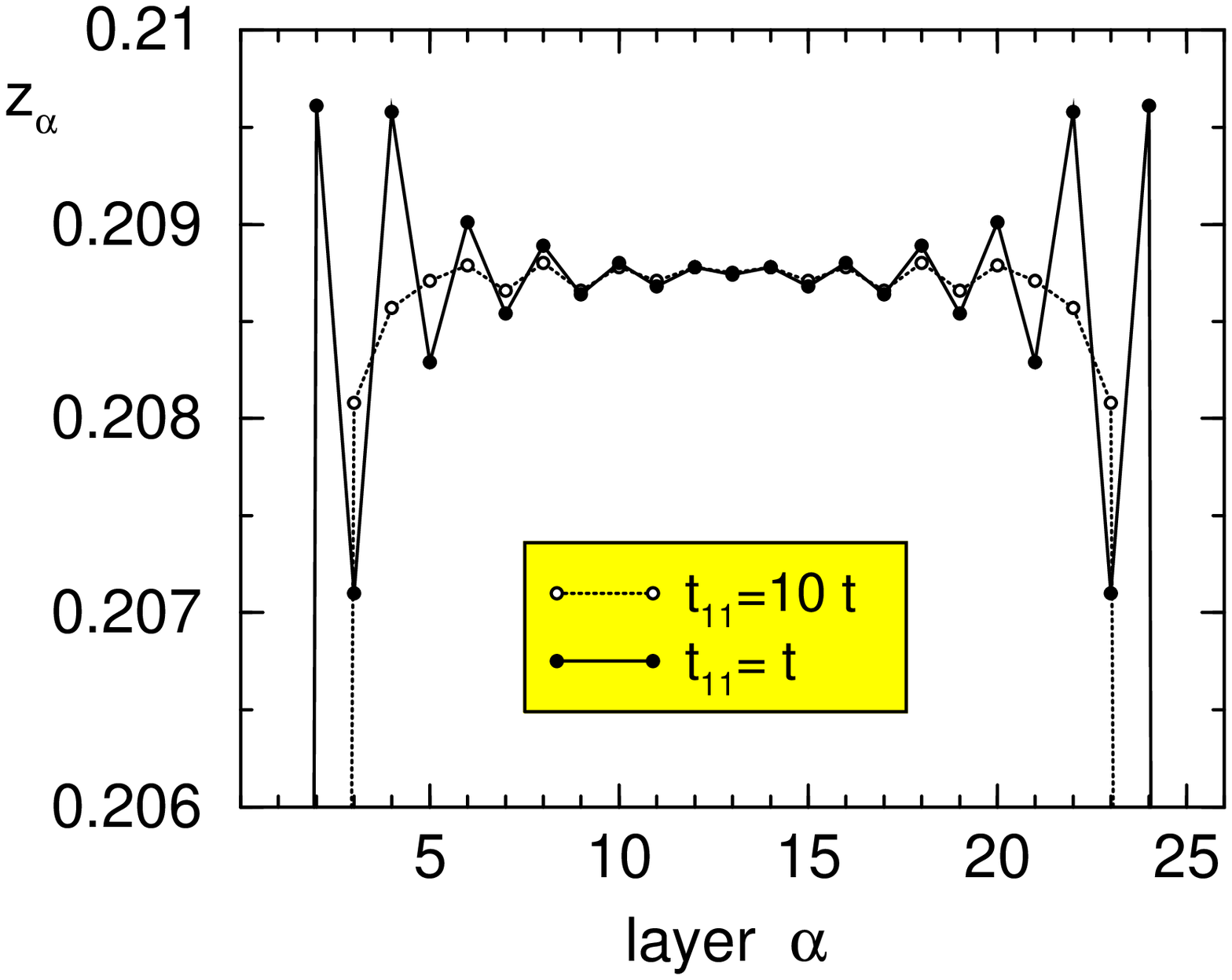,width=80mm,angle=0}}
\vspace{2mm}

\begin{center}
\parbox[]{120mm}{\small Fig.\ \ref{fig:para3}:
The same as Fig.~\ref{fig:para4} but on an enlarged scale. 
Slab thickness: $d=25$.
\label{fig:para3}
}
\end{center}
\end{figure}

\refstepcounter{figure}
\begin{figure}[ht] 
\vspace{4mm}
\centerline{\psfig{figure=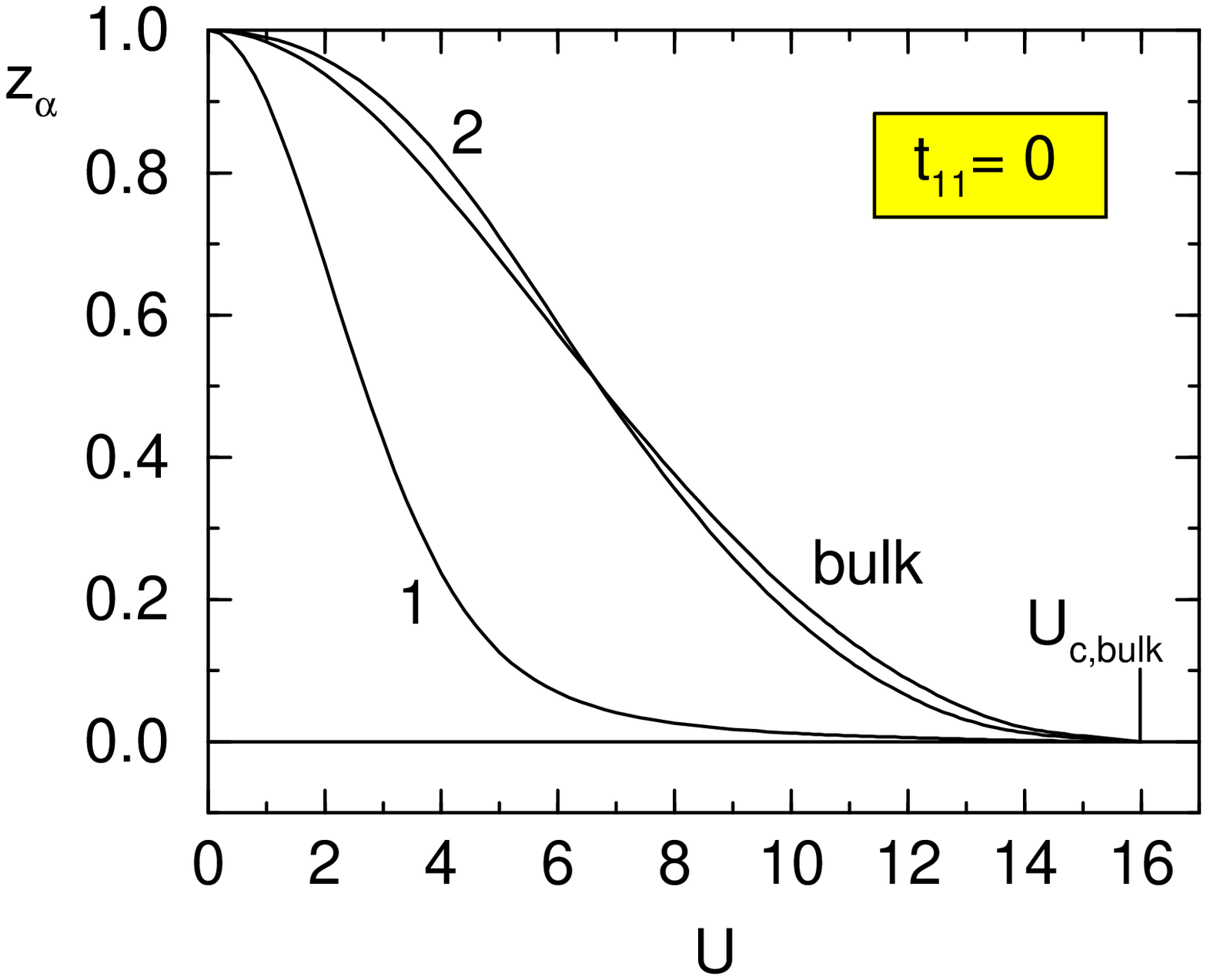,width=80mm,angle=0}}
\vspace{2mm}

\begin{center}
\parbox[]{120mm}{\small Fig.\ \ref{fig:t11zero}:
$U$ dependence of $z_\alpha$ when the intra-layer surface hopping
is switched off.
\label{fig:t11zero}
}
\end{center}
\end{figure}

A modification of the model parameters at the very surface may
strongly affect the quasi-particle weight. As in Sec.\ \ref{sec:s5} 
we first
consider a modified hopping within the top layer: $t_{11} \ne t$. 

Fig.~\ref{fig:para1} 
gives an overview for fixed Coulomb interaction $U=10$.
The above-mentioned tendency towards an insulating surface is 
enhanced when $t_{11}$ is decreased. The top-layer quasi-particle
weight quickly decreases but even for $t_{11}=0$ it does not vanish 
completely. For $t_{11} > t$ one can see the opposite trend.
$z_{\alpha=1}$ increases with increasing $t_{11}$. In the limit
$t_{11} \mapsto \infty$ it approaches its non-interacting value
$z_{\alpha=1}=1$. For $t_{11}=10 t$ the low-energy 
electronic structure is 
almost perfectly decoupled. In the top surface layer there is a 
quasi uncorrelated motion of the electrons ($z_{\alpha=1} = 0.98$
at $U/t_{11}=1$). The rest system, however, remains to be a strongly 
correlated Fermi liquid.

For the sub-surface layers, the dependence of the quasi-particle 
weight on $t_{11}$ is comparatively weak. Fig.\ \ref{fig:para2} 
shows $z_\alpha$
for $\alpha \ge 2$. On the enlarged scale in Fig.\ \ref{fig:para2} 
there is still 
a considerable $t_{11}$ dependence of $z_{\alpha=2}$ (second layer).
For $\alpha \mapsto \infty$, however, i.~e.\ with increasing distance 
to the surface, this dependence diminishes: The bulk quasi-particle 
weight obviously cannot be affected by the surface modification of 
the hopping parameter. We also notice that there is a nearly constant 
quasi-particle weight for $t_{11} \approx 0.8 t$ and all
$\alpha \ge 3$.

For fixed $t_{11}$ one finds an oscillating layer dependence of 
$z_\alpha$. This is demonstrated in Figs.~\ref{fig:para4} and
\ref{fig:para3} for $t_{11} = t$ 
and $t_{11} = 10t$. For the strongly perturbed system with 
$t_{11}=10$, the layer dependence is somehow irregular in the 
surface-near region, oscillations do not build up until $\alpha \ge 5$. 
In both cases the oscillation is strongly damped. For $\alpha=13$ 
we have $\Delta z / z \approx 2 \cdot 10^{-4}$. Thus, for a 
film with thickness $d=25$, the quasi-particle weight is nearly 
constant at the film center. Furthermore, the differences between the 
uniform and the perturbed system become smaller and smaller with 
increasing distance to the surface. The observed oscillations can 
be traced back to oscillations of the free layer DOS 
at the Fermi energy. It is well known \cite{HK73} that the presence 
of the surface gives rise to a layer-by-layer oscillation of 
$\rho_\alpha(E=0)$ for $U=0$. For the present case (local 
self-energy, manifest particle-hole symmetry, metallic phase), 
the density of states at the Fermi edge is unrenormalized by the 
interaction (see Eq.\ (\ref{eq:onsitegf}) and Ref.\ \cite{PN99a}). 
The same oscillation is thus found for $\rho_\alpha(E=0)$ at any
$U<U_{c, {\rm bulk}}$ and will also lead to oscillations of the
low-energy part of the Green function and thereby to oscillations
of the low-energy part of the self-energy. Finally, the oscillating 
behavior
of $z_\alpha$ for $U=10$ shows that we are well below the critical
point: For $U$ close to $U_{c, {\rm bulk}}$ we expect a monotonous 
behavior from the linearized DMFT (see Fig.\ \ref{fig:prof}).

Let us now tackle the question of surface phases. The scenario of
an insulating surface coexisting with a metallic bulk was excluded 
by the linearized DMFT. The same is found by the numerical 
evaluation of the DMFT: Fig.~\ref{fig:t11zero} 
shows the 
layer-dependent quasi-particle weight for $t_{11}=0$ where the
strongest suppression of $z_{\alpha=1}$ is expected. In fact, 
the top-layer quasi-particle weight quickly decreases as a 
function of $U$ and, compared with the bulk value, becomes very
small above $U\approx 6$. However, we find a non-zero weight in
the top layer up to $U=U_{c,{\rm bulk}}$ which implies 
$U_{c,{\rm surf}} = U_{c,{\rm bulk}}$. Between $U\approx 6$ and
$U=U_{c,{\rm bulk}}$ we may speak of an {\em induced} metallic
surface according to the discussion in Sec.\ \ref{sec:s5}.

The linearized DMFT predicted a metallic surface on top of a 
Mott-insulating bulk to be possible for $t_{11} > t \sqrt{5/4}$.
We choose $t_{11} = 1.5 t$ for the numerical calculation to be
well above this threshold. Fig.~\ref{fig:trans1} 
proves that two different
critical interactions are found indeed. Over the whole $U$ range
considered, the top-layer quasi-particle weight is strongly
enhanced compared with the bulk and is finite also at
$U=U_{c,{\rm bulk}}$ where the bulk weight vanishes.
Note that $z_{\alpha=1}(U)$ is continuous at $U=U_{c,{\rm bulk}}$. 
The top-layer quasi-particle weight 
approaches zero at $U=U_{c,{\rm surf}} = 20.0$ 
which marks the surface transition point while the extraordinary 
transition takes place at $U=U_{c,{\rm bulk}}=16.0$.

\refstepcounter{figure}
\begin{figure}[t] 
\vspace{4mm}
\centerline{\psfig{figure=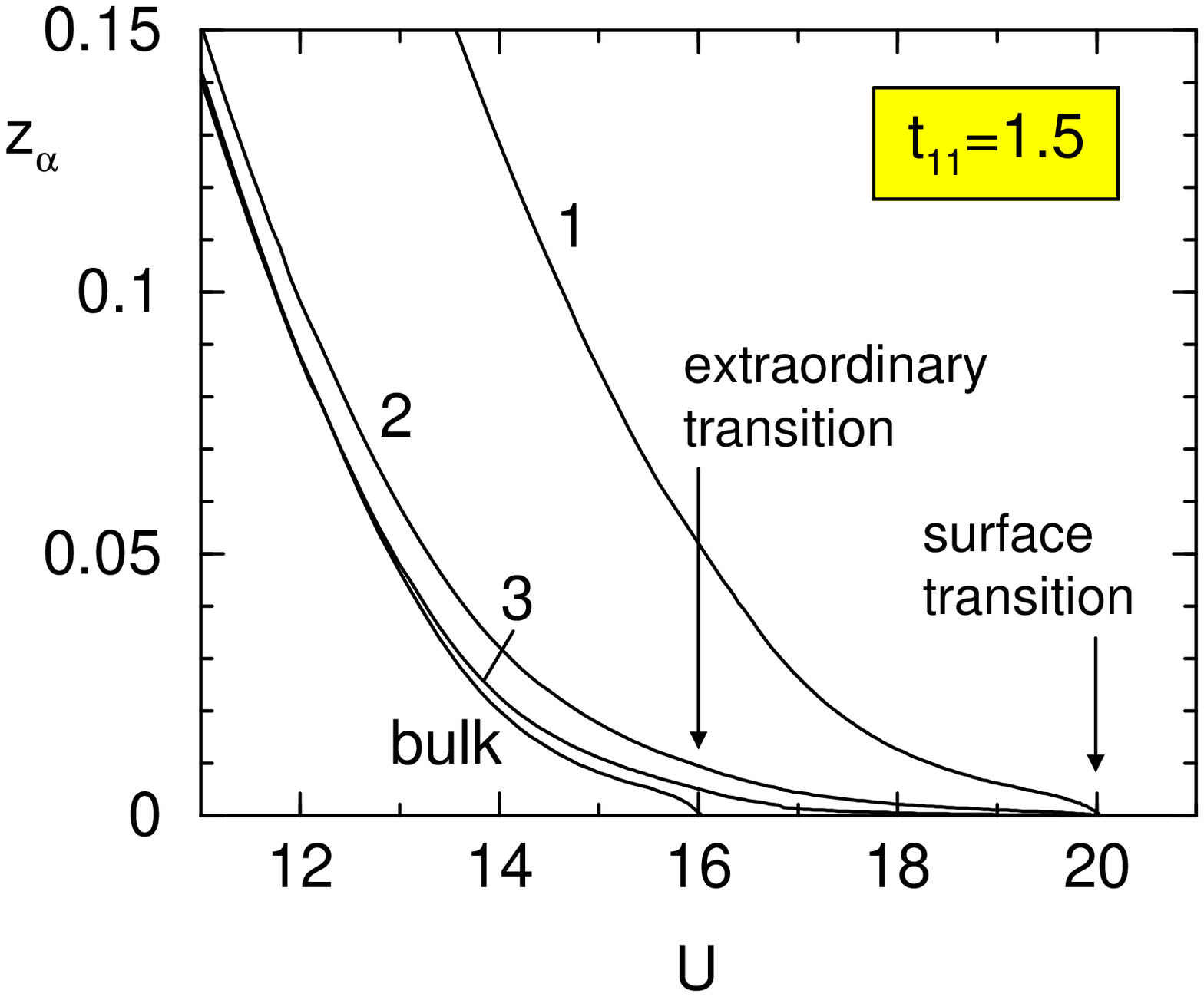,width=80mm,angle=0}}
\vspace{2mm}

\begin{center}
\parbox[]{120mm}{\small Fig.\ \ref{fig:trans1}:
$U$ dependence of $z_\alpha$ for enhanced intra-layer surface
hopping. Surface transition at $U=U_{c, {\rm surf}}$. Extraordinary
transition at $U=U_{c, {\rm bulk}}$.
\label{fig:trans1}
}
\end{center}
\end{figure}

\refstepcounter{figure}
\begin{figure}[t] 
\vspace{4mm}
\centerline{\psfig{figure=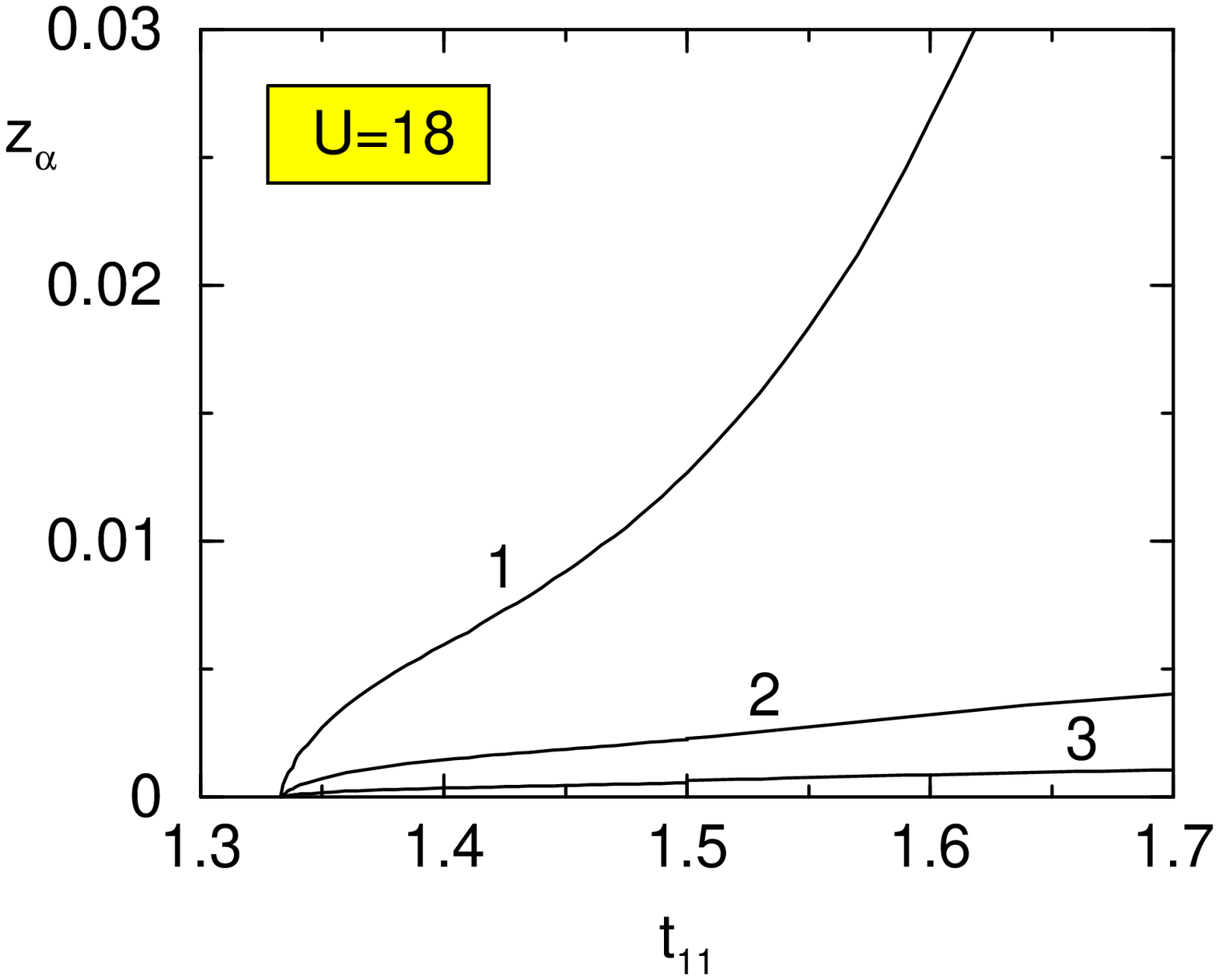,width=80mm,angle=0}}
\vspace{2mm}

\begin{center}
\parbox[]{120mm}{\small Fig.\ \ref{fig:para5}:
$z_\alpha$ as a function of $t_{11}$ for $U>U_{c,{\rm bulk}}$.
\label{fig:para5}
}
\end{center}
\end{figure}

Evaluating the analytical formula for the surface critical 
interaction (\ref{eq:ucpara}) for the present case, we get
$U_{c,{\rm surf}} = 18.2$ which agrees well with the
numerical result if one takes into account that also for the
bulk critical interaction the linearized theory yields a 
somewhat smaller value. We also expect that $U_{c,{\rm surf}}$
(as $U_{c,{\rm bulk}}$) is overestimated by the ED due to 
finite-size effects \cite{PN99a,PN99c}. While finite-size effects 
prevent a precise determination of the critical interactions, 
they are irrelevant concerning the very existence of the metallic 
surface phase. Even for $U$ well above $U_{c,{\rm bulk}}$, the 
top-layer quasi-particle weight is still larger than 
$z_{\alpha=1}=0.01$, and thus the ED for $n_s=8$ is still able 
to resolve the energy scale set by the width of the Kondo-type 
resonance at the surface.

Since the low-energy surface excitations cannot propagate into 
the bulk for $U>U_{c,{\rm bulk}}$ but are reflected at the bulk
Hubbard gap, the Kondo resonance represents a true surface state.
Therefore, its amplitude must decay exponentially with increasing 
distance to the surface. Fig.~\ref{fig:trans1} 
shows that some weight is induced 
in the sub-surface layers which rapidly decreases.

The surface transition is also found as a function of $t_{11}$
for fixed interaction $U > U_{c,{\rm bulk}}$. Fig.~\ref{fig:para5} 
shows the
numerical results for $U=18$. When $t_{11}>t_{11,{\rm c}}=1.33t$
the surface becomes metallic. The critical value may be compared
with $t_{11,{\rm c}}=1.48t$ which is obtained by solving Eqs.\ 
(\ref{eq:cond}) and (\ref{eq:surfev}) for $t_{11}$.

\refstepcounter{figure}
\begin{figure}[b] 
\vspace{4mm}
\centerline{\psfig{figure=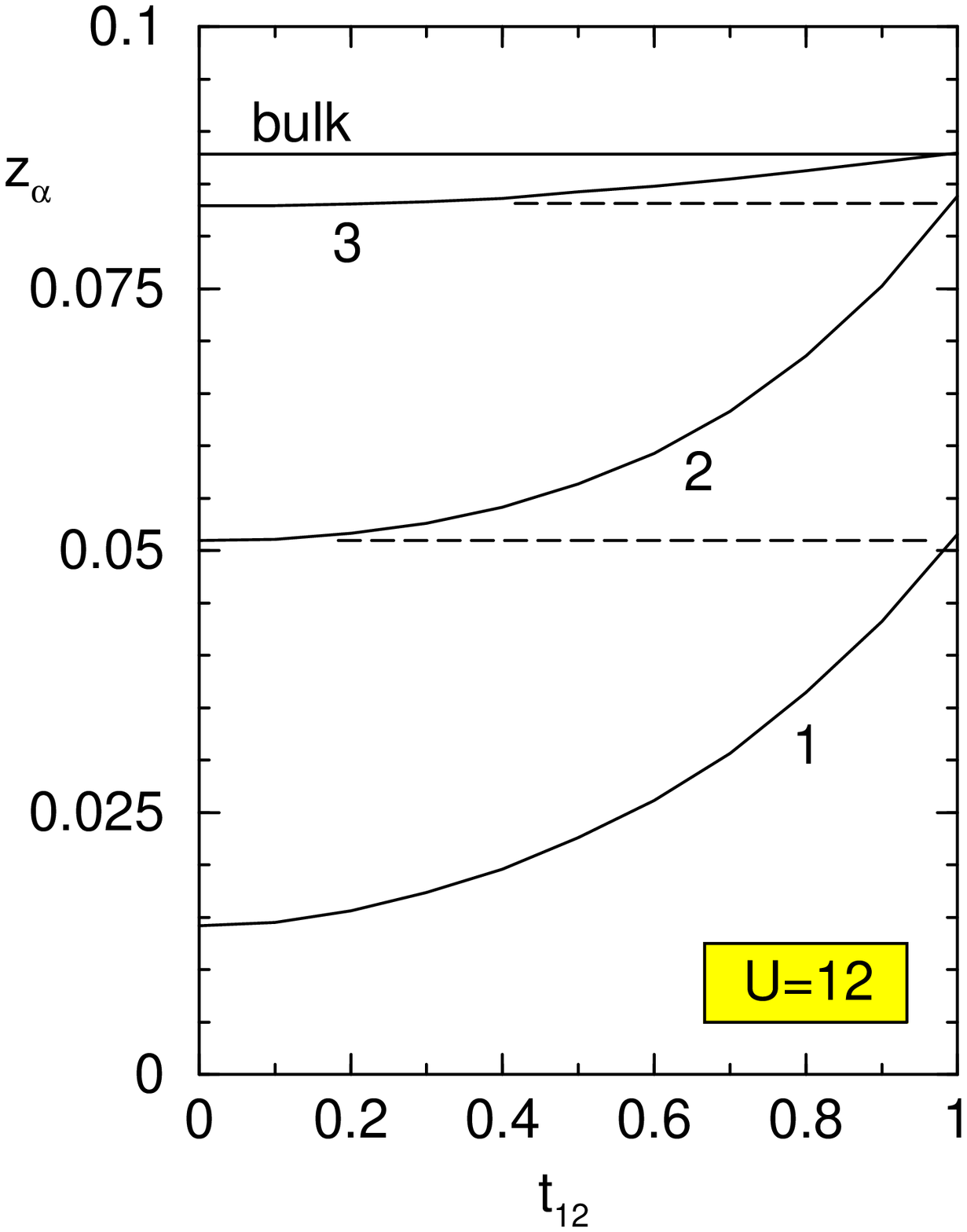,width=80mm,angle=0}}
\vspace{2mm}

\begin{center}
\parbox[]{120mm}{\small Fig.\ \ref{fig:perp1}:
Layer-dependent quasi-particle weight for $U<U_{c,{\rm bulk}}$
as a function of the inter-layer surface hopping $t_{12} \le t =1$.
\label{fig:perp1}
}
\end{center}
\end{figure}

For the $t_{11}$ range considered in Fig.~\ref{fig:para5}, 
the number of layers 
in the slab $d$ that is necessary to simulate the actual surface 
can be lowered down to $d \approx 5$: We performed calculations for 
different thicknesses $d$; there are hardly any differences between 
the results for $z_\alpha$ at the surface as long as $d \ge 5$. 
This is interpreted as follows: Since the coherent part of the bulk 
spectrum has disappeared for $U > U_{c,{\rm bulk}}$, the surface 
electronic structure is essentially decoupled from the bulk in
the low-energy regime. The decoupling at the low-energy scale is 
indicated by the rapid decrease of $z_\alpha$ with increasing 
$\alpha$ (see Fig.~\ref{fig:para5}). 
Contrary, on the high-energy scale set by 
the charge excitations, bulk and surface modes cannot decouple. There 
is always a finite energetic overlap of the bulk and the surface 
DOS since $t_{11}$ mainly changes the effective widths but not 
the positions of the Hubbard peaks in the surface DOS. The effect 
of the Hubbard bands on the low-energy features, however, seems 
to be rather weak since otherwise a change of $d$ would lead to 
significant changes in the surface low-energy electronic structure 
by indirect coupling between low- and high-energy surface excitations 
and high-energy surface and bulk excitations. The surface Kondo 
resonance in the metallic surface phase is spatially confined to
the first few layers and energetically isolated from the surface
Hubbard bands.

\subsection{Modified inter-layer surface hopping} 

\refstepcounter{figure}
\begin{figure}[b] 
\vspace{4mm}
\centerline{\psfig{figure=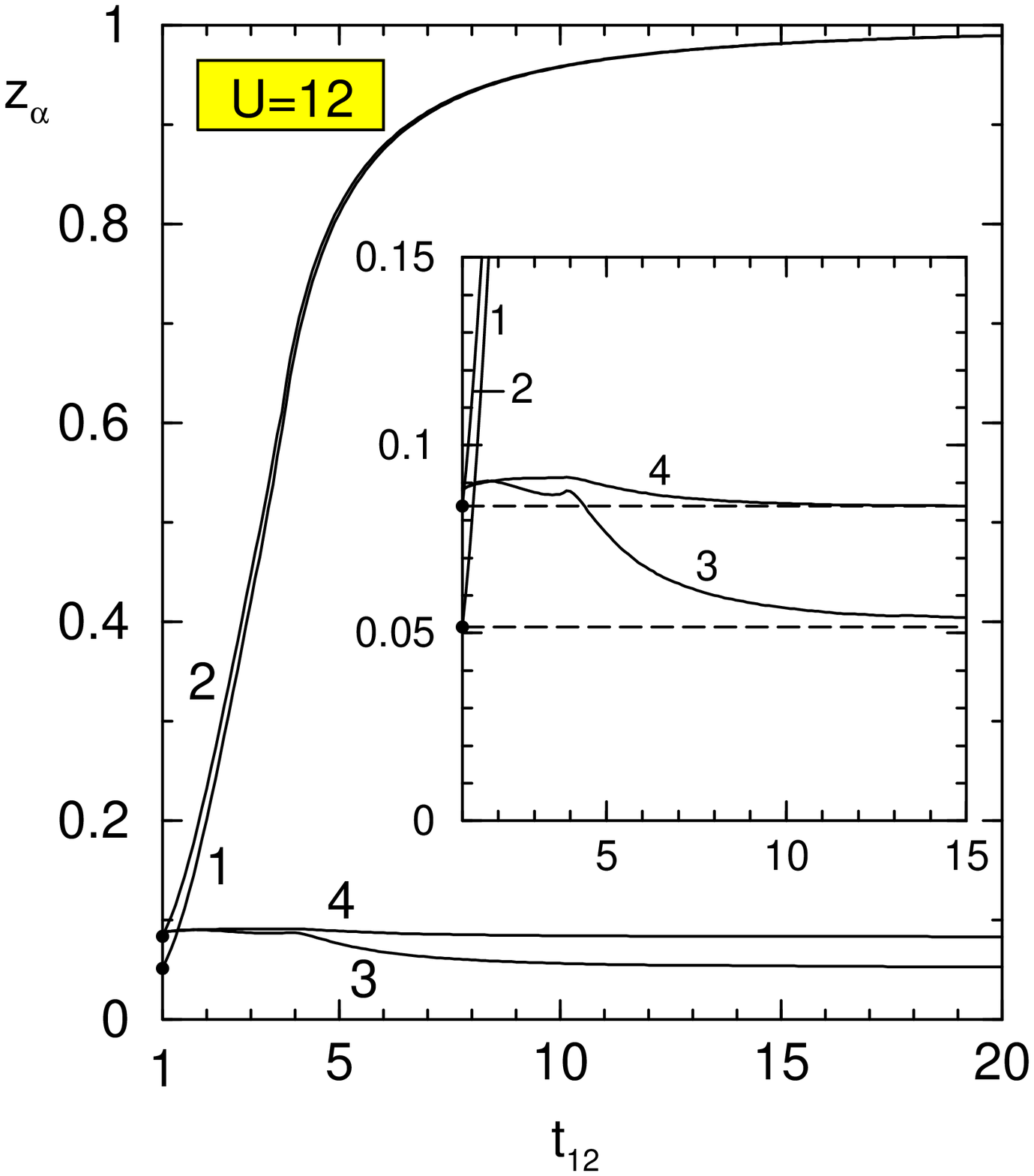,width=80mm,angle=0}}
\vspace{2mm}

\begin{center}
\parbox[]{120mm}{\small Fig.\ \ref{fig:perp2}:
The same as Fig.~\ref{fig:perp1} 
but for $t_{12} \ge t = 1$. Inset: asymptotic
behavior of $z_\alpha$.
\label{fig:perp2}
}
\end{center}
\end{figure}

\refstepcounter{figure}
\begin{figure}[t] 
\vspace{4mm}
\centerline{\psfig{figure=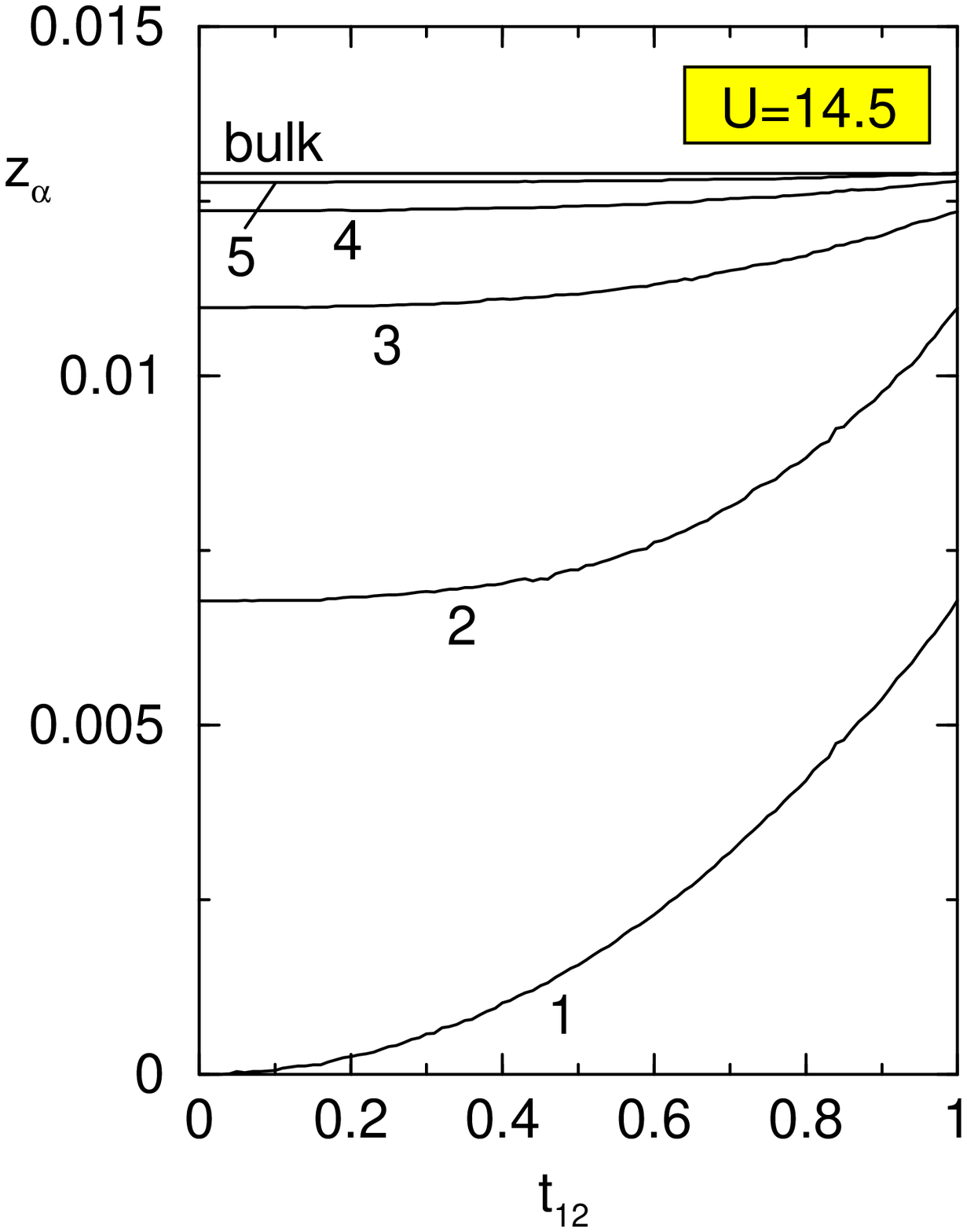,width=60mm,angle=0}}
\vspace{2mm}

\begin{center}
\parbox[]{120mm}{\small Fig.\ \ref{fig:perp2d3d}:
The same as Fig.~\ref{fig:perp1} 
but for $U_{c, D=2} < U < U_{c,{\rm bulk}}$.
\label{fig:perp2d3d}
}
\end{center}
\end{figure}

\refstepcounter{figure}
\begin{figure}[ht] 
\vspace{4mm}
\centerline{\psfig{figure=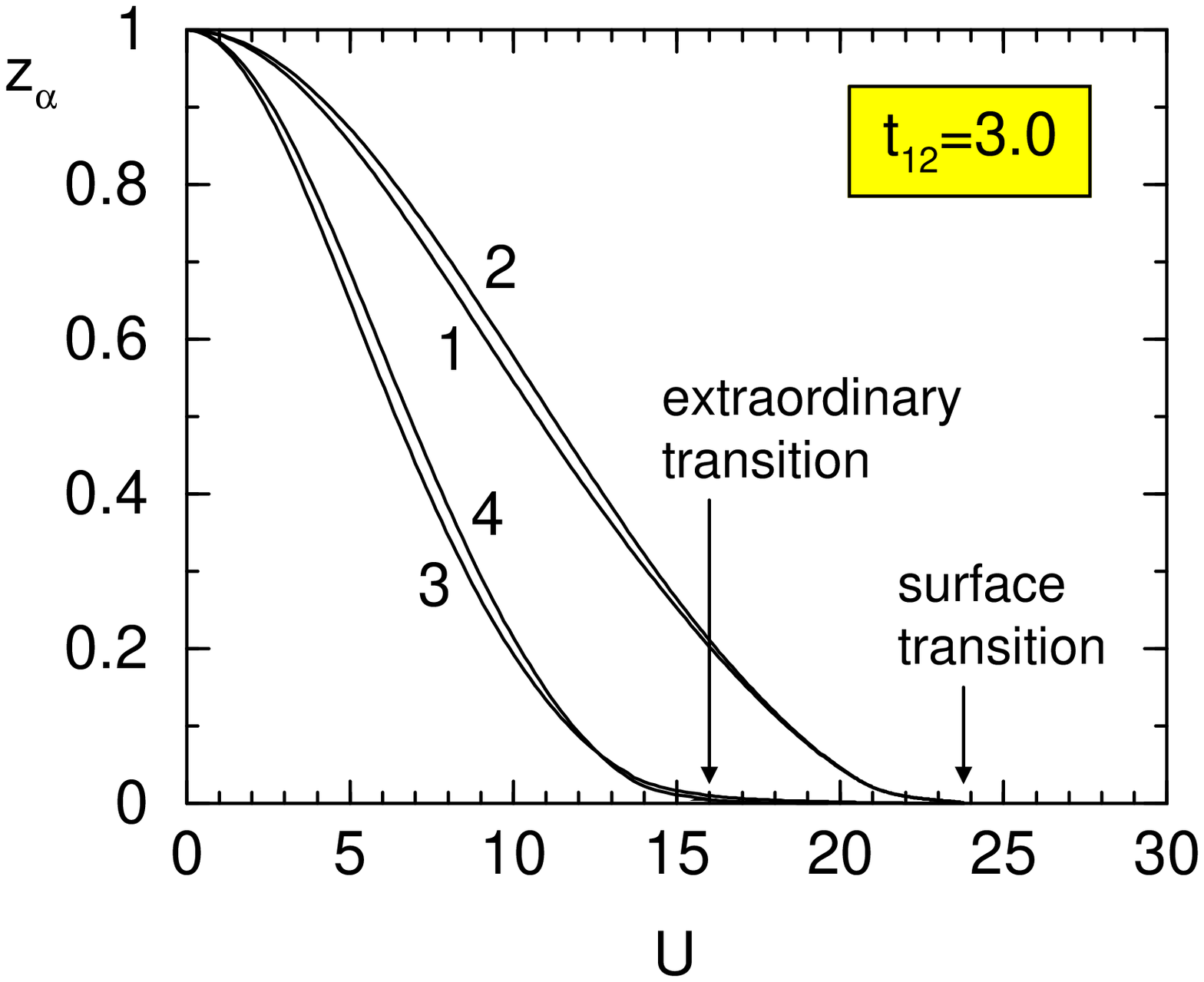,width=70mm,angle=0}}
\vspace{2mm}

\begin{center}
\parbox[]{120mm}{\small Fig.\ \ref{fig:trans2}:
$U$ dependence of $z_\alpha$ for enhanced inter-layer surface
hopping $t_{12}$.
\label{fig:trans2}
}
\end{center}
\end{figure}

A complete decoupling between the top layer and the rest system
is obtained for vanishing inter-layer surface hopping $t_{12}=0$.
Fig.~\ref{fig:perp1} 
shows the layer-dependent quasi-particle weight as a 
function of $t_{12}$. While for $U=10$ we noticed an oscillating
layer dependence for uniform hopping (Figs.~\ref{fig:para4} and
\ref{fig:para3}), there is 
a monotonous layer dependence for $U=12$ (Fig.~\ref{fig:perp1}, 
for $t_{12}=1$). 
This is the typical behavior when the system is close to criticality
as has been noted before (cf.\ Ref.\ \cite{PN99c} and the discussion 
of the analytical $z_\alpha$ profiles in Sec.\ \ref{sec:s5}). 
The layer dependence remains
to be monotonous for $t_{12} \mapsto 0$. For $t_{12} = 0$ we have
essentially two independent systems. The isolated top layer is 
still metallic. In the rest system the $\alpha=2$ layer represents 
the new top layer, the $\alpha=3$ layer becomes the first sub-surface 
layer, and so on. This implies that the value of $z_\alpha$ for 
$t_{12}=0$ must be equal to the value of $z_{\alpha-1}$ for 
$t_{12}=1$. These relations are indicated by the dashed lines
in Fig.~\ref{fig:perp1}. 
They represent a non-trivial check of the numerics.

An effective separation into subsystems is also observed in the 
opposite limit of a strongly enhanced inter-layer surface hopping.
Fig.~\ref{fig:perp2} 
shows that $z_{\alpha=1}$ and $z_{\alpha=2}$ approach
their non-interacting values while for $\alpha \ge 3$ the 
quasi-particle weight changes only slightly as $t_{12}\mapsto\infty$.
In the low-energy regime the electronic structure of the first
two layers decouples from the rest system. The value of 
$z_{\alpha}$ for all $\alpha \ge 3$ approaches the value
of $z_{\alpha-2}$ for $t_{12}=1$ (see inset).

A somewhat artificial realization of an insulating surface phase
on top of a metallic bulk can be obtained for $t_{12} \mapsto 0$
by choosing $U_{c, D=2} < U < U_{c,{\rm bulk}}$, where $U_{c,D=2}$
is the critical interaction of the two-dimensional layer. For
$U > U_{c, D=2}$ the top layer must become insulating when it is
decoupled from the rest system ($t_{12}=0$). This is demonstrated
in Fig.~\ref{fig:perp2d3d}. 
The figure also shows that the top layer becomes 
metallic (with a very small quasi-particle weight) as soon as
an arbitrarily small inter-layer hopping is switched on.

Finally, Fig.~\ref{fig:trans2} 
shows the extraordinary and the surface transition
for fixed $t_{12}=3.0$. The surface critical interaction can be
read off to be $U_{c,{\rm surf}} = 23.8$ while the linearized DMFT
with $U_{c,{\rm surf}}=21.7$ [Eq.\ (\ref{eq:ucperp})] again predicts
a slightly smaller critical value.

\subsection{Modified surface Coulomb interaction} 

\refstepcounter{figure}
\begin{figure}[b] 
\vspace{4mm}
\centerline{\psfig{figure=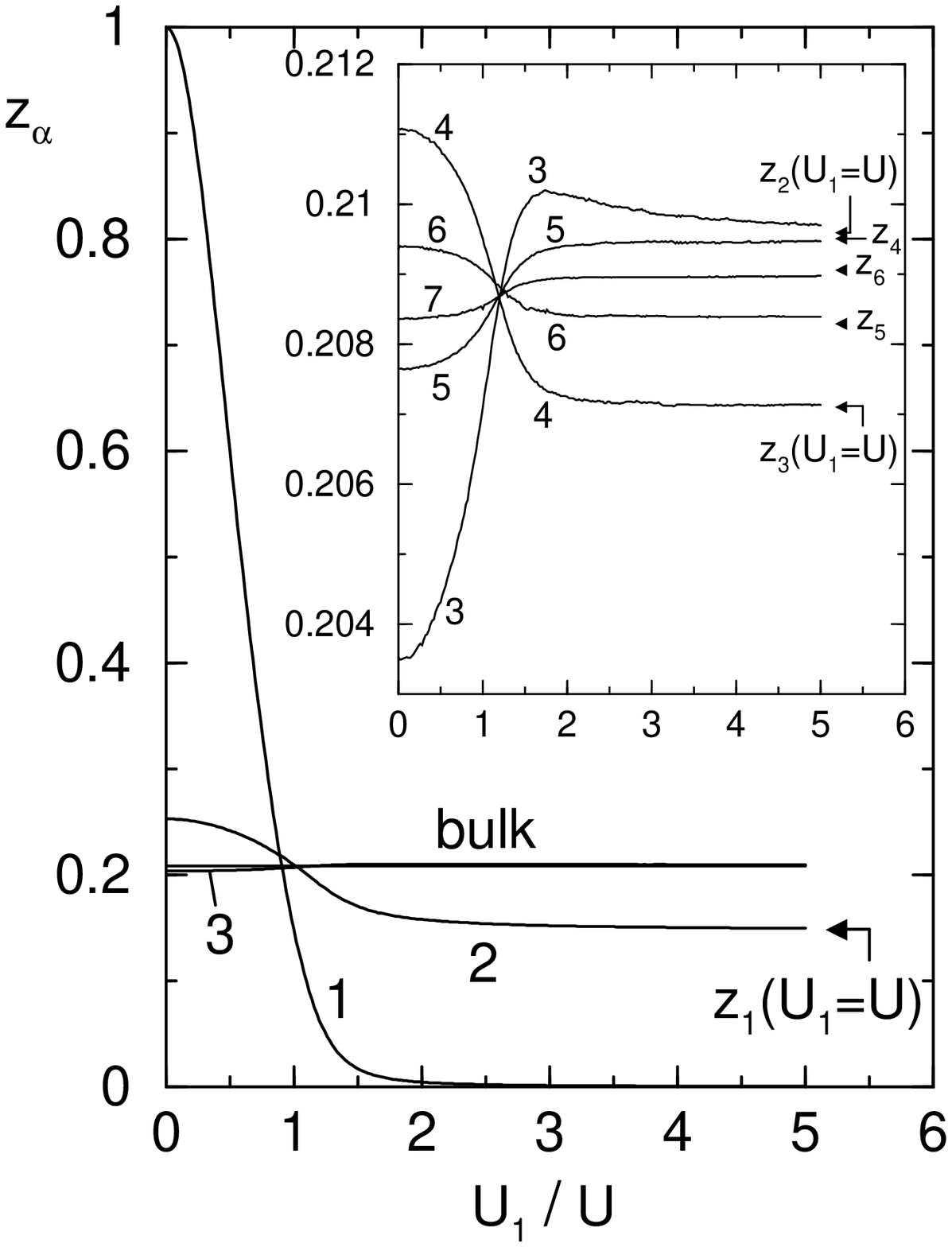,width=80mm,angle=0}}
\vspace{2mm}

\begin{center}
\parbox[]{120mm}{\small Fig.\ \ref{fig:ux}:
Layer-dependent quasi-particle weight for $U=10<U_{c,{\rm bulk}}$
and modified Coulomb interaction in the top layer $U_1$. The arrow
indicates the value of $z_{\alpha=1}$ for $U_1=U$. In the inset
the arrows indicate the $U_1=U$ values of $z_\alpha$.
\label{fig:ux}
}
\end{center}
\end{figure}

\refstepcounter{figure}
\begin{figure}[t] 
\vspace{4mm}
\centerline{\psfig{figure=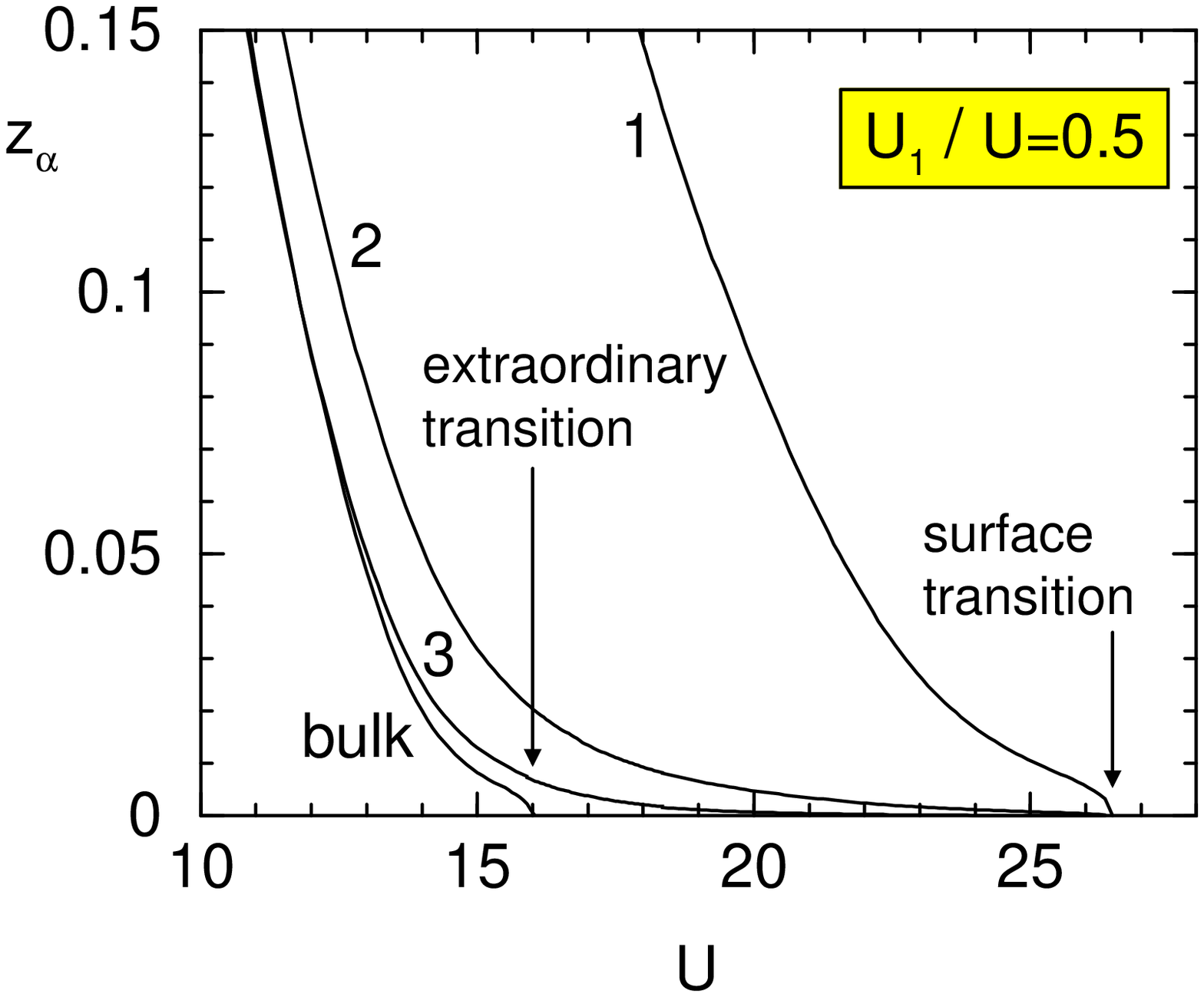,width=80mm,angle=0}}
\vspace{2mm}

\begin{center}
\parbox[]{120mm}{\small Fig.\ \ref{fig:trans3}:
$U$ dependence of $z_\alpha$ for enhanced surface Coulomb 
interaction $U_1/U={\rm const.}$
\label{fig:trans3}
}
\end{center}
\end{figure}

We finally discuss the modification of the surface Coulomb interaction
$U_1$. Fig.\ \ref{fig:ux} shows the quasi-particle weight $z_\alpha$ 
for $\alpha=1,2,3$
and $z_{\rm bulk}$ as a function of $U_1/U$ where $U$ is fixed at
$U=10$. On decreasing $U_1$ ($U_1 < U$), $z_1$ quickly increases, and
for $U_1 \mapsto 0$ it approaches the non-interacting value $z_1=1$.

For enhanced $U_1 > U$ the top-layer quasi-particle weight is
decreased but remains to be finite even for large large values of
$U_1$, i.~e.\ we again find an {\em induced} metallic surface.
Asymptotically, however, the top-layer weight approaches zero:
$z_1 \mapsto 0$ for $U_1 \mapsto \infty$. In this limit the 
low-energy resonance in the top-layer DOS essentially disappears
and a large Hubbard gap $\sim U_1$ opens. This implies that
-- in the low-energy regime -- the sub-surface ($\alpha=2$) DOS
for $U_1\mapsto \infty$ must become identical with the $U_1=U$
surface ($\alpha=1$) DOS. For $U_1\mapsto \infty$ we thus expect
$z_2(U_1\mapsto \infty) = z_1(U_1=U)$ and consequently
$z_\alpha(U_1\mapsto \infty) = z_{\alpha-1}(U_1=U)$ for all $\alpha$.
In fact, this ``shift'' of the surface by one layer can be seen in
Fig.\ \ref{fig:ux} and in the inset: For $U_1/U=5$ only small 
differences still remain between $z_\alpha$ and $z_{\alpha-1}(U_1=U)$.

The ``shift'' $\alpha\mapsto \alpha-1$ also implies that the
oscillating layer dependence of $z_\alpha$ for $U_1=U$ must be
reversed for $U_1\mapsto \infty$: Minima are replaced by maxima
and vice versa. This partly explains that between (at 
$U_1/U\approx 1.2$) the quasi-particle weight is nearly layer
independent.

According to the linearized DMFT, a metallic surface on top of
an insulating bulk can be found if $U_1 < 6t \sqrt{5}$ 
[Eq.\ (\ref{eq:ccuu})] and $U>U_{c, {\rm bulk}} = 6t \sqrt{6}$.
If we fix the ratio $U_1/U=0.5$ and vary $U$, the surface 
transition should occur at $U_{1, c, {\rm surf}} = 12.1$
(set $U=2U_{1, c, {\rm surf}}$ in Eq.\ (\ref{eq:mm3}) and solve
for $U_{1, c, {\rm surf}}$). The result of the numerical 
solution of the DMFT equations is shown in Fig.\ \ref{fig:trans3}.
Again, the numerically obtained value, 
$U_{1, c, {\rm surf}} = 0.5 \cdot 26.3 = 13.15$, is somewhat larger
than the analytical prediction -- the discussion is the same as
for the modified surface hopping.

At the extraordinary transition $U=U_{c, {\rm bulk}}$, the
quasi-particle weight in the top- 
(not shown in Fig.\ \ref{fig:trans3}) and in the first
sub-surface layers are smooth functions of $U$. This is 
contrary to the results found within the slave-boson theory
\cite{Has92} where the band-narrowing factor in the top layer
shows up a discontinuous derivative 
at $U=U_{c, {\rm bulk}}$. We believe,
however, that this is an artifact due to incorrect boundary
conditions. Namely, only the first two surface layers are treated
as ``free'' in the self-consistent calculation while already the
$\alpha=3$ layer is assumed to be bulk-like in Ref.\ \cite{Has92}.
As is known from the mean-field theory of localized spin models
\cite{MLS89}, such boundary conditions may result in artificial
singularities at the extraordinary transition.

\section{Conclusion} 
\label{sec:s8}

We have investigated the (Mott) metal-insulator transition at 
surfaces within the framework of the semi-infinite Hubbard model
at half-filling and $T=0$.
Basically, two approximations have been used:

First, the self-energy functional has been assumed to be reasonably 
local. This approximation sets the basis for the dynamical mean-field 
theory: The semi-infinite Hubbard model is self-consistently mapped 
onto a set of indirectly coupled impurity models corresponding
to the inequivalent layers parallel to the surface. With the usual
scaling of the (intra-layer and inter-layer) hopping, the approach 
becomes exact in the limit of infinite spatial dimensions. 
It has been shown that there are non-trivial surface effects even
for $D=\infty$. Mainly, however, the DMFT has been used as a
(mean-field) approach to study the $D=3$ low-index surfaces 
of the simple-cubic lattice.

Second, for the approximate solution of the impurity models, we have
used the exact diagonalization of finite systems. The ED method allows 
to deal systematically with a large number of geometries and model 
parameters. However, the method cannot access the very critical 
regime for the Mott transition because of errors due to finite-size 
effects. Directly at the critical point we have alternatively 
considered a 
simplification of the mean-field equations (linearized DMFT).
This analytical approach is also approximate. However, a convincing 
qualitative and (as far as can be judged) also quantitative 
agreement with the numerical ED results has been found. 

Referring to the points mentioned in the introduction, our
results can be summarized as follows:

\begin{enumerate}
\item 
The metal-insulator transition in the bulk of the semi-infinite 
system occurs exactly at the same critical interaction 
$U_{c,{\rm bulk}}$ as for the infinitely extended system: 
$U_{c,{\rm bulk}}=U_c$.
\item
There is a non-trivial layer dependence of the quasi-particle weight,
even (asymptotically) at the critical point. The $z_\alpha$ profile
strongly depends on the model parameters at the surface, e.~g.\
the hopping within the top layer $t_{11}$, the hopping between the
top and the sub-surface layer $t_{12}$ and the top-layer Coulomb
interaction $U_1$. There is a qualitative change of the profile
if certain critical values, $t_{11,c}$, $t_{12,c}$ and $U_{1,c}$,
are exceeded. These critical values are found to be of a realistic
order of magnitude.
\item
For uniform model parameters the top-layer quasi-particle weight 
$z_1$ is smaller than the bulk value $z_{\rm bulk}$ since a reduced 
surface coordination number implies correlation effects to be 
effectively stronger at the surface. For interactions well below
$U_c$, there is always an oscillating layer dependence of the
the quasi-particle weight. With increasing distance to the surface
($\alpha \mapsto \infty$), this oscillation is strongly damped. 
In the critical regime, on the contrary, $z_\alpha$ monotonously 
increases with increasing $\alpha$, and finally, for $U=U_c$ the 
critical profile is linear: $z_\alpha \propto \alpha$.
For uniform model parameters there is a finite weight 
in the top layer ($z_1>0$) for $U<U_c$ only, i.~e.\ only when 
the bulk is metallic. The transition at $U_c$ is termed the
``ordinary transition''.
\item
For a sufficiently strong modification of the surface model 
parameters ($t_{11}>t_{11,c}$, $t_{12}>t_{12,c}$, $U_1<U_{1,c}$),
the surface becomes metallic below a critical interaction
$U_{c,{\rm surf}} > U_{c,{\rm bulk}}$ (``surface transition''). 
For $U_{c,{\rm bulk}} < U < U_{c,{\rm surf}}$, 
the quasi-particle weight  
exponentially decays from its maximum value $z_1$ at the 
surface towards zero in the bulk. At $U=U_{c,{\rm bulk}}$ 
the bulk undergoes the transition to the metallic state
(``extraordinary transition''). The top-layer quasi-particle 
weight is a smooth function of $U$ even at $U=U_{c,{\rm bulk}}$.
\item
The transition at $U = U_{c,{\rm surf}} = U_{c,{\rm bulk}}$ 
is called the ``special transition''. Here the critical profile
of the quasi-particle weight is flat 
$z_\alpha=z_{\rm bulk}=\mbox{const.}$ (at least for $\alpha\ge 2$).
In this situation the effect of missing neighbors at the surface 
is compensated by the change of the surface model parameters.
\item
There are two critical exponents that are merely related to 
the critical interactions, the 
``shift exponent'' $\lambda_s$ and the ``crossover exponent''
$\phi$. They describe the trend of the $U_c$ for a film of finite
thickness $d$ in the limit $d\mapsto \infty$ and the trend of the
surface critical interaction for the semi-infinite system near 
the special transition, respectively.
Within the linearized DMFT one finds $\lambda_s=2$ and $\phi=1/2$.
\item
For any realistic choice of the model parameters, a metallic bulk 
induces a metallic surface with $z_1>0$. Thus, a Mott-insulating 
surface of a correlated metal is impossible.
There are essentially two more or less trivial exceptions: The first
is the static decoupling of the top layer for $t_{12}=0$ at an
interaction strength that is smaller than $U_{c,{\rm bulk}}$ but
larger than the critical interaction of the two-dimensional system.
The second is a dynamical decoupling which occurs for 
infinite surface interaction $U_1 \mapsto \infty$. Here the
top-layer quasi-particle weight vanishes asymptotically.
\end{enumerate}

\section*{Acknowledgement} 

This work is supported by the Deutsche Forschungsgemeinschaft 
within the SFB 290.

\end{document}